\documentclass[pdftex,12pt]{article}

\usepackage{amsmath}
\usepackage{amsfonts}
\usepackage{amssymb}
\usepackage{amsthm}
\usepackage{physics}
\usepackage{color}
\usepackage{appendix}
\usepackage{graphicx}

\usepackage{geometry} 
\usepackage{cite}
\usepackage[utf8]{inputenc}
\usepackage{mcite}

\usepackage[colorlinks=true,linkcolor=blue,citecolor=magenta,linktocpage=true]{hyperref}

\usepackage{bbm} 
\newcommand{\one}{\mathbbm{1}}

\newcommand{\pr}[1]{\left(#1\right)}
\newcommand{\cor}[1]{\left[#1\right]}
\newcommand{\cur}[1]{\left\{#1\right\}}
\newcommand{\oyleki}{\,\vert\,}

\newcommand{\R}{{\mathbb{R}}}
\newcommand{\Z}{{\mathbb{Z}}}
\newcommand{\N}{{\mathbb{N}}}
\newcommand{\C}{\mathbb{C}}
\newcommand{\pd}{\partial}
\newcommand{\half}{{\frac12}}

\newcommand{\tell}{\tilde{\ell}}
\newcommand{\tambda}{{\tilde{\lambda}}}
\newcommand{\tx}{{\tilde{x}}}
\newcommand{\ty}{{\tilde{y}}}
\newcommand{\tk}{{\tilde{k}}}
\newcommand{\tn}{{\tilde{n}}}

\newcommand{\Hambda}{{\bar{\Lambda}}}

\newcommand{\X}{{\mathbb{X}}}
\newcommand{\Y}{{\mathbb{Y}}}
\newcommand{\K}{{\mathbb{K}}}
\newcommand{\QQ}{{\hat{\mathbb{Q}}}}
\newcommand{\Qq}{{\mathbb{Q}}}
\newcommand{\AAA}{{\mathbb{A}}}
\newcommand{\dX}{{\dot{\mathbb{X}}}}
\newcommand{\ddX}{{\ddot{\mathbb{X}}}}
\newcommand{\WW}{{\mathbb{W}}}
\newcommand{\PP}{{\mathbb{P}}}

\newcommand{\WS}{{\mathcal{W}}}
\newcommand{\WL}{{\mathcal{W}_\Lambda}}
\newcommand{\PSp}{{\mathcal{P}}}
\newcommand{\TL}{{T_\Lambda}}
\newcommand{\ML}{{M_\Lambda}}
\newcommand{\EL}{E_\Lambda}

\newcommand{\Lag}{\mathcal{L}}
\newcommand{\Ham}{\mathcal{H}}
\newcommand{\HOp}{\hat{H}}
\newcommand{\HG}{G}
\newcommand{\DS}{\mathcal{D}}
\newcommand{\Ord}{\mathcal{O}}

\newcommand{\Weyl}[2]{{\hat{W}_{(#1,#2)}}}
\newcommand{\Weylp}[2]{{\hat{W}_{(#1,#2)}^{\phantom{\dagger}}}}
\newcommand{\Weyld}[2]{{\hat{W}_{(#1,#2)}^{\dagger}}}
\newcommand{\Wey}[1]{{\hat{W}_{#1}}}

\newcommand{\ketS}[1]{\ket{#1}_{\mathrm{Sch}}}
\newcommand{\ketM}[1]{\ket{#1}_{\mathrm{mom}}}
\newcommand{\ketL}[1]{\ket{#1}_\Lambda}
\newcommand{\keto}[1]{\vert #1 \rangle}

\newcommand{\alphaS}{\alpha_{\mathrm{Sch}}}
\newcommand{\alphaM}{\alpha_{\mathrm{mom}}}

\newcommand{\subeq}[1]{\begin{subequations}#1\end{subequations}}

\newcommand{\mtrx}[1]{{\renewcommand{\arraystretch}{1.0}\begin{pmatrix}#1\end{pmatrix}}}

\newtheorem{lem}{Lemma}

\begin{document}

\begin{titlepage}

\title{Path Integral in Modular Space}
\author{Yigit Yargic$^{1,2}$ \thanks{yyargic@perimeterinstitute.ca}}
\date{\small{\textit{
$^1$Perimeter Institute for Theoretical Physics,\\ 31 Caroline Street North, Waterloo, Ontario, N2L 2Y5, Canada \\[5pt]
$^2$Department of Physics and Astronomy, University of Waterloo,\\ 200 University Avenue West, Waterloo, Ontario, N2L 3G1, Canada\\}}}

\maketitle

\begin{abstract}

The modular spaces are a family of polarizations of the Hilbert space that are based on Aharonov's modular variables and carry a rich geometric structure. We construct here, step by step, a Feynman path integral for the quantum harmonic oscillator in a modular polarization. This \emph{modular path integral} is endowed with novel features such as a new action, winding modes, and an Aharonov-Bohm phase. Its saddle points are sequences of superposition states and they carry a non-classical concept of locality in alignment with the understanding of quantum reference frames. The action found in the modular path integral can be understood as living on a compact phase space and it possesses a new set of symmetries. Finally, we propose a prescription analogous to the Legendre transform, which can be applied generally to the Hamiltonian of a variety of physical systems to produce similar modular actions.

\end{abstract}

\end{titlepage}

\tableofcontents



\section{Introduction}

The notion of space that underlies the physical reality has been a crucial element for most approaches in the history of physics. In Newtonian mechanics, this is a three-dimensional space, which describes the possible positions of a given object. We call this space here the \emph{Schr\"odinger space} after the corresponding polarization in quantum mechanics. Since the phase space variables do not commute as quantum operators, the representations in quantum theory require choosing a commutative subset of these variables, which also defines an underlying space as a basis. Although the Schr\"odinger space carries a special role for its link to the classical descriptions, it is essentially one among many polarizations of the Hilbert space.

The correspondence principle posits that the calculations in quantum mechanics must reproduce classical calculations in the limit of large quantum numbers, or $\hbar \rightarrow 0$. Although this requirement has been historically useful for the development of quantum theory, it is widely accepted today that quantum physics is a more fundamental and accurate description of Nature, while classical physics is merely an approximation thereof. The reliance on the Schr\"odinger space is a classical relic in quantum mechanics -- one we aim to eliminate in this paper.

It is natural to ask here what we can replace the Schr\"odinger space with. It is advocated in the study of quantum reference frames \cite{Giacomini:2017zju} that polarizations are observer-dependent properties and do not carry a fundamental meaning for Nature. In this respect, any quantum reference frame can provide a consistent description for physics. However, since each representation carries a different geometry with various properties, it will be advantageous for us to formulate the physics on a background that is both simple and rich in its geometric structures.

A generic class of polarizations of the Weyl algebra known as the \emph{modular representations} \cite{Freidel:2016pls} will be our main point of focus in this paper. In fact, both Schr\"odinger and momentum representations are included in this class as two opposite limits, as we will discuss. The idea for the modular representations goes back to the study of modular variables \cite{Aharonov:1969qx,aharonov_2003} for the Aharonov-Bohm effect.

A modular representation is based on the quotient of the phase space by a so-called \emph{modular lattice}. This quotient is a torus with the volume $\pr{2\pi\hbar}^d$ and it is called a \emph{modular space}. Since the modular space has twice the number of dimensions as the Schr\"odinger space, a modular state is labeled by a pair of position and momentum variables. Although it is counter-intuitive to be able to label a quantum state with both its position and momentum, these labels are defined periodically with respect to the modular lattice, therefore they reconcile with Heisenberg's uncertainty principle.

The modular spaces have several advantages over the Schr\"odinger space. Firstly, they admit a length and momentum scale in their construction, while also respecting Lorentz symmetry \cite{Freidel:2016pls}. This observation makes them valuable in the pursuit for quantum gravity where a fundamental scale, the Planck scale, has to be incorporated into the quantum structure of spacetime without breaking its symmetries. Secondly, the limits in which the Schr\"odinger and momentum representations are obtained from the modular ones are singular, and many geometric structures, such as the symplectic structure and an Abelian gauge symmetry, are lost from the configuration space in this limit. This carries the risk that any approach to quantum gravity that relies on the classical space can be missing these essential geometric ingredients. Finally, the modular representations form a continuous class. This makes it possible for future research to explore the physical consequences of infinitesimal changes in the quantum reference frame and therefore in the notion of locality.

Our work in this paper is focused on replacing the Schr\"odinger space with the modular space in one particular framework of the quantum theory: the Feynman path integral. In the standard formulation of the Feynman path integral \cite{Feynman:1948ur}, a quantum transition amplitude is expressed as a weighted sum over all trajectories between two points in the Schr\"odinger space. Each one of these trajectories is a sequence of classical configurations, i.e.~local states in the Schr\"odinger space. Therefore, the Feynman path integral only seems to support the historical misconception that the Schr\"odinger space has a preferred status in quantum mechanics and justify its use at the foundation of quantum gravity.

In this paper, we give a step-by-step construction of a path integral in the modular space, following analogous steps to Feynman's original construction. We use the Hamiltonian of a quantum harmonic oscillator in the framework of non-relativistic quantum mechanics for this purpose. The resulting \emph{modular path integral} is a weighted sum over trajectories in the modular space. Each trajectory in this path integral, including its saddle points, is a sequence of classically non-local states. As this non-locality is manifested in quantum superpositions, one may interpret the trajectories in the path integral as being purely quantum mechanical. Another example of a path integral with this particular feature has been studied in \cite{Green:2016gpb} using tensor network techniques. The modular path integral provides further evidence that the quantum theory does not \emph{need} the classical notion of Schr\"odinger space and breaks the associated concept of locality.

As expected, our modular path integral displays some novel features that vanish in the singular Schr\"odinger limit. Firstly, it contains a new action on the phase space with a larger set of symmetries. Secondly, since the modular space is toroidally compact, the path integral involves a sum over the winding number around the modular space. Thirdly, the trajectories are weighted by an additional phase that depends on their winding number, which is analogous to an Aharonov-Bohm phase.

There are two possible approaches to argue for the role of modular spaces in fundamental physics. In the ``radical'' approach, one may postulate that the underlying geometry of Nature is a modular space(-time) with a preferred scale. This proposal arose from the study of Born reciprocity \cite{Freidel:2013zga} and metastring theory \cite{Freidel:2015pka}, and it has been further pursued by Freidel, Leigh and Minic in \cite{Freidel:2016pls,Freidel:2017wst,Freidel:2017yuv,Freidel:2018apz}. In particular, a path integral in the phase space is sketched in \cite{Freidel:2018apz} from a different perspective, where it is interpreted as representing the trajectories of a so-called metaparticle.

The second, ``algebraic'' approach to the modular spaces is to consider them on the same footing as the Schr\"odinger space as arbitrary polarizations of the Weyl algebra. In this approach, the Nature does not have a commutative space(-time) for a preferred background, but each polarization serves as a reference frame aligned to an observer or a physical subsystem. For example, the quantum state of an electron that passes through an infinite grid is projected onto a modular state. This electron sees itself as localized in a modular space (whose length scale is given by the spacing of the grid), while a lab observer sees the electron in a non-local superposition. This example demonstrates that locality can be an observer-dependent property in quantum mechanics. This approach is aligned with the perspective brought by the study of quantum reference frames \cite{Giacomini:2017zju,Vanrietvelde:2018pgb}. We remain agnostic between these two approaches in this paper.

Since the action found in the modular path integral differs from the standard action, it is interesting to formulate the transformation from a Hamiltonian to the associated modular Lagrangian as a new kind of Legendre transform. We conjecture here a new prescription, called the \emph{modular Legendre transform}, which can be applied more generally to a wider range of physical systems to obtain a similar modular action.

The paper is organized as follows. In Section \ref{sec:Representations}, we introduce the modular representation of the Weyl algebra and discuss its relationship with the Schr\"odinger representation. We give the explicit construction of a path integral in modular space in Section \ref{sec:PathIntegralConstruction} using the Hamiltonian for a quantum harmonic oscillator. The equation \eqref{MPI-Final} marks our main result in this paper. We analyze the new modular action in Section \ref{sec:Analysis} for its solutions and canonical formulation. In Section \ref{sec:LimitSch}, we discuss how the modular path integral recovers the Feynman path integral in a certain limit. We conclude in Section \ref{sec:Conclusion} with the discussion and interpretation of our results.

\section{Representations of the Weyl algebra}
\label{sec:Representations}

We consider the quantum mechanics of a single non-relativistic particle in $d$ dimensions. The position and momentum operators, $\hat{q}^a$ and $\hat{p}_a$, $a=1,...,d$, obey Heisenberg's canonical commutation relation $[\hat{q}^a,\hat{p}_b] = i\hbar \, \delta^a_b$.

Since the position and momentum operators are unbounded, it is advantageous to consider their exponentiated versions. For each $a,b \in \R^d$, where $a$ has the units of length and $b$ has the units of momentum, we define the \emph{Weyl operator} $\Weyl{a}{b}$ by
\begin{align}
    \Weyl{a}{b} &\equiv
	e^{i (b \cdot \hat{q} - a \cdot \hat{p}) / \hbar}
	\;.
\end{align}
The Weyl operators build the \emph{Weyl algebra} $\mathcal{W}$ together with the relations
\begin{align}
    \Weyld{a}{b}
    &=	\Weylp{-a}{-b}
    \;, \\
    \Weyl{a}{b} \, \Weyl{a'}{b'}
    &=	e^{\half i ( b \cdot a' - a \cdot b' ) / \hbar} \,
	    \Weyl{a + a'}{b + b'}
    \;.
\end{align}
As such, the Weyl algebra is a non-commutative C*-algebra.

In order to construct a representation of the Weyl algebra $\mathcal{W}$, we usually choose a commutative subalgebra of $\mathcal{W}$ that becomes diagonalized in this representation. Once a commutative C*-subalgebra is chosen, the Gelfand-Naimark theorem \cite{GelfandNaimark} provides an associated topological space, such that the subalgebra is isometrically *-isomorphic to an algebra of complex functions on this space. We view this space provided by the Gelfand-Naimark theorem as the \emph{quantum configuration space} for the chosen representation of the Weyl algebra.

In the following, we will briefly review the standard Schr\"odinger representation and its dual to set up our notation, then we will introduce the modular representations, which are the focus of this paper.

\subsection{Schr\"odinger representation}

The Schr\"odinger representation is based on a commutative subalgebra of $\mathcal{W}$ that is spanned by the elements $\{ \Weyl{0}{b} \oyleki b \in \R^d \}$. In other words, the position operators $\hat{q}^a$ and their exponentials are diagonalized in this representation. Their common eigenvectors are denoted by $\ketS{x}$, $x \in \R^d$, and they satisfy $\hat{q}^a \ketS{x} = x^a \ketS{x}$ and $\Weyl{0}{b} \ketS{x} = e^{i b \cdot x / \hbar} \ketS{x}$.

A general quantum state can be written in the Schr\"odinger representation as
\begin{align}
    \ket{\psi} =
    \int_{\R^d} \dd^d x \;
    \psi(x) \ketS{x}
    \;,
\end{align}
where $\psi \in L^2(\R^d)$ is the Schr\"odinger wave function. The momentum operators act on the wave functions as $\hat{p}_a \psi(x) \sim - i \hbar \, \frac{\pd}{\pd x^a} \psi(x)$.

One can similarly construct the momentum representation from the commutative subalgebra of $\mathcal{W}$ spanned by $\{ \Weyl{a}{0} \oyleki a \in \R^d \}$. The momentum eigenvectors $\ketM{\tx}$, $\tx \in \R^d$, satisfy $\hat{p}_a \ketM{\tx} = \tx_a \ketM{\tx}$, and they are related to the position eigenvectors by a Fourier transform, ${}^{\phantom{*}}_{\mathrm{Sch}}\!\braket{x}{\tx}_{\mathrm{mom}} = \pr{2\pi\hbar}^{-d/2} e^{i x \cdot \tx / \hbar}$.

\subsection{Modular representation}
\label{sec:ModRepr}

After having considered two standard examples, we may now look for the generic commutative subalgebras of the Weyl algebra $\mathcal{W}$. The commutator of two Weyl operators can be written as
\begin{align}
    \cor{ \Weyl{a}{b} , \Weyl{a'}{b'} } &=
    \pr{ e^{i ( b \cdot a' - a \cdot b' ) / \hbar} - 1 }
    e^{- \half i ( b \cdot a' - a \cdot b' ) / \hbar} \,
	\Weyl{a + a'}{b + b'}
    \;.
\end{align}
This implies that
\begin{align}
\label{ModLat-SolveCom2}
    \cor{ \Weyl{a}{b} , \Weyl{a'}{b'} } = 0
    \qquad \Leftrightarrow \qquad
    \frac{1}{2\pi\hbar} \pr{a' \cdot b - a \cdot b'}
    \in \Z
    \;.
\end{align}
Since this relation is bilinear, the arguments $(a,b) \in \R^{2d}$ of the Weyl operators in a generic commutative subalgebra of $\mathcal{W}$ are supported on a lattice in the phase space.

Hereafter, we follow the notation in \cite{Freidel:2016pls}, where double-stroke, capital letters such as $\X^A = (x^a, \tx_a)$ denote a pair of position and momentum variables, which are represented by lowercase letters without and with a tilde, respectively. These composite objects are vectors on the phase space $\PSp = \R^d \oplus \R^d$.

We introduce a symplectic structure $\omega$ on $\PSp$ by $\omega(\X,\Y) = \omega_{AB} \, \X^A \, \Y^B \equiv \tx \cdot y - x \cdot \ty$ for any $\X, \Y \in \PSp$. We say that $\Lambda \subset \PSp$ is a \emph{modular lattice} if it is a maximal subset that satisfies $\omega(\Lambda,\Lambda) \subseteq 2\pi\hbar \, \Z$. Our discussion above shows that each maximal commutative *-subalgebra $\WL \subset \WS$ corresponds to a modular lattice $\Lambda$, i.e.~it is generated by the elements $\{ \Wey{\K} \oyleki \K \in \Lambda \}$.\footnote{The Schr\"odinger and momentum representations correspond to two singular limits of modular lattices, which we will discuss later in Section \ref{sec:SingularLimits}.}

In the following, we will assume that the modular lattice $\Lambda$ is of the form $\Lambda = \{ (\lambda n, \tambda \tn) \in \PSp \oyleki n,\tn \in \Z^d \}$, where $\lambda^a{}_b$ and $\tambda_a{}^b$ are diagonal $d \times d$-matrices, which satisfy $\lambda^c{}_a \tambda_c{}^b = 2\pi\hbar \, \delta_a^b$. Note that any modular lattice can be brought to this form by a symplectic coordinate transformation.

The common eigenvectors of $\WL$ are called the \emph{modular vectors}. A modular vector $\ketL{\X}$, $\X \in \PSp$, can be expressed in terms of Schr\"odinger's position eigenvectors through a Zak transform \cite{Zak01}, such that\footnote{\label{Footnote-ZakMom}These modular vectors can equivalently be expressed in terms of momentum eigenvectors as
\begin{align}
\label{ModVec-DefMom}
    \ketL{\X} &\equiv
    ( \det\lambda )^{-1/2} \,
    e^{- \half i x \cdot \tx / \hbar}
    \sum_{\tn \in \Z^d} e^{-i x \cdot \tambda \tn / \hbar}
    \, \keto{\tx + \tambda \tn}_{\mathrm{mom}}^{\phantom{\dagger}}
    \;.
\end{align}}
\begin{align}
\label{ModVec-DefSch}
    \ketL{\X} &\equiv
    \big( \! \det\tambda \big)^{-1/2} \,
    e^{\half i x \cdot \tx / \hbar}
    \sum_{n \in \Z^d} e^{i \tx \cdot \lambda n / \hbar}
    \ketS{x + \lambda n}
    \;.
\end{align}
These vectors satisfy the eigenvalue equation
\begin{align}
\label{ModVec-Eigen}
    \Wey{\K} \ketL{\X} &=
    e^{\half i k \cdot \tk / \hbar} \,
    e^{i \omega(\K,\X) / \hbar}
    \ketL{\X}
    \;, \qquad
    \K \in \Lambda \;,\;
    \X \in \PSp
    \;.
\end{align}
for any 
The action of a generic Weyl operator on a modular vector is given by
\begin{align}
\label{ModVec-WeylActs}
    \Wey{\Y} \ketL{\X} &=
    e^{\half i \omega(\Y,\X) / \hbar}
    \ketL{\X + \Y}
    \;, \qquad
    \X, \Y \in \PSp
    \;.
\end{align}
Moreover, the modular vectors are quasi-periodic under discrete translations along the modular lattice, such that
\begin{align}
\label{ModVec-QuasiPer}
    \ketL{\X + \K} &=
    e^{\half i k \cdot \tk / \hbar} \,
    e^{\half i \omega(\K,\X) / \hbar} \ketL{\X}
    \;, \qquad
    \K \in \Lambda \;,\; \X \in \PSp
    \;.
\end{align}
In the following, we will often drop the subscript $\Lambda$ on modular vectors for better readability.

The quasi-periodicity implies that not all modular vectors are linearly independent. In order to construct a basis from the modular vectors, we consider the quotient $T_\Lambda \equiv \PSp / \Lambda$, which is called a \emph{modular space}. Each element\footnote{We abuse the notation by using the same symbol $\X$ both for the elements of $\PSp$ as well as for the corresponding equivalence classes on $\TL$.} $\X \in \TL$ of the modular space is an equivalence class of the points $(\X + \Lambda) \subset \PSp$, which can be represented by any of those points. The modular space is topologically a torus in $2d$ dimensions and it has the volume $(2\pi\hbar)^d$. It is the associated Gelfand-Naimark space for a modular representation, which we will regard as a quantum configuration space.

We define a \emph{modular cell} $\ML \subset \PSp$ as any set of representatives of the modular space $\TL$. Then, the vectors $\{ \ketL{\X} \! \oyleki \X \in \ML \}$ form a complete and orthonormal basis of the Hilbert space. The orthogonality relation reads
\begin{align}
\label{ModVec-Orthogonality}
    \braket{\X}{\Y} = \delta^{2d}(\X - \Y)
    \;, \qquad \X, \Y \in \ML
    \;,
\end{align}
where $\delta^{2d}$ denotes the $2d$-dimensional Dirac delta distribution. While the relation \eqref{ModVec-Orthogonality} gives the inner product of two modular vectors from the same modular cell, the inner product of two generic modular vectors is given by
\begin{align}
\label{ModVec-Inner}
    \braket{\X}{\Y} = \sum_{\K \in \Lambda}
    e^{\half i k \cdot \tk / \hbar} \,
    e^{\half i \omega(\K,\X) / \hbar} \,
    \delta^{2d}(\X - \Y + \K)
    \;, \qquad \X, \Y \in \PSp
    \;.
\end{align}
The completeness relation for modular vectors reads
\begin{align}
\label{ModVec-Completeness}
    \one =
    \int_{\TL} \dd^{2d}\X \, \ketbra{\X}{\X}
    \;.
\end{align}
Note that writing the integration in \eqref{ModVec-Completeness} over the modular space $\TL$ employs the fact that $\ketbra{\X}{\X}$ is periodic and therefore independent of the choice of the modular cell.

We can write a general quantum state in the modular basis as
\begin{align}
    \ket{\phi} =
    \int_{\TL} \dd^{2d}\X \; \phi(\X) \ketL{\X}
    \;,
\end{align}
where $\phi(\X)$ is called a \emph{modular wave function}\footnote{This function can be thought of as mapping $\phi : \PSp \rightarrow \C$ under the restriction \eqref{ModWaveF-QuasiPer}, while a more rigorous definition is given below in terms of $\EL$.}. This integral is well-defined only when the integrand $\phi(\X) \ketL{\X}$ is periodic on $T_\Lambda$. Therefore, we require the modular wave functions to be also quasi-periodic, such that
\begin{align}
\label{ModWaveF-QuasiPer}
    \phi(\X + \K) &=
    e^{- \half i k \cdot \tk / \hbar} \,
    e^{- \half i \omega(\K,\X) / \hbar} \,
    \phi(\X)
    \;, \qquad
    \K \in \Lambda \;,\; \X \in \PSp
    \;.
\end{align}
In order to reformulate this statement in a more abstract way as in \cite{Freidel:2016pls}, one may define a $U(1)$-bundle $\EL \rightarrow \TL$ over the modular space together with the identification
\begin{align}
    \EL : \quad
    \pr{\theta,\X} \sim
    \pr{\theta \, 
        e^{\half i k \cdot \tk / \hbar} \,
        e^{\half i \omega(\K,\X) / \hbar} ,
        \X + \K }
    \;, \quad
    \K \in \Lambda \;,\; \X \in \PSp \;,\; \theta \in U(1)
    \;.
\end{align}
Then, the modular wave functions $\phi \in L^2(\EL)$ correspond to the square-integrable sections of $\EL$.

Finally, we examine the action of Heisenberg operators $\hat{q}^a$ and $\hat{p}_a$ on a quantum state in the modular representation. After some calculation, we find
\subeq{\begin{align}
    \hat{q}^a \ket{\phi} &=
    \int_{\TL} \dd^{2d}\X \pr{
        i\hbar \, \frac{\pd}{\pd \tx_a} \, \phi(\X)
        + \half \, x^a \, \phi(\X)
        } \ketL{\X}
    \;, \\
    \hat{p}_a \ket{\phi} &=
    \int_{\TL} \dd^{2d}\X \pr{
        - i\hbar \, \frac{\pd}{\pd x^a} \, \phi(\X)
        + \half \, \tx_a \, \phi(\X)
        } \ketL{\X}
    \;.
\end{align}}
These equations can be expressed more compactly in terms of an Abelian connection\footnote{Unlike all other double-stroke letters in this paper, $\AAA$ denotes a co-vector on $\PSp$, rather than a vector.} $\AAA = \AAA_A \, \dd\X^A$ on $\EL$, given by
\begin{align}
    \AAA_A(\X) \equiv \pr{\half \, \tx_a , - \half \, x^a}
    \;.
\end{align}
The key property of this \emph{modular connection} $\AAA$ is that its curvature form coincides with the symplectic form, i.e.~$\dd\AAA = \omega$. Using the modular connection, we can define a covariant derivative $\nabla$, which acts on the modular wave functions as
\begin{align}
    \nabla_A \phi(\X) &\equiv
    \pd_A \phi(\X) + \frac{i}{\hbar} \, \AAA_A(\X) \, \phi(\X)
    \;,
\end{align}
where $\pd_A \equiv \big( \frac{\pd}{\pd x^a} , \frac{\pd}{\pd \tx_a} \big)$. Defining also $\QQ^A \equiv (\hat{q}^a,\hat{p}_a)$, we can finally write the action of Heisenberg operators in the modular representation as $\QQ^A \sim i\hbar \, (\omega^{-1})^{AB} \, \nabla_B$.

One can check that the actions of the Weyl operators $\Wey{\Y}$ and the Heisenberg operators $\QQ^A$ on a modular wave function preserve the condition \eqref{ModWaveF-QuasiPer}, therefore these are well-defined operators on the modular Hilbert space $L^2(\EL)$.

\subsection{Modular gauge transformation}

There is a $U(1)$-gauge freedom in defining the modular vectors as follows. For any real, smooth function $\alpha \in C^\infty(\PSp)$ on the phase space, we may redefine the modular vectors as
\begin{align}
\label{ModGau-Transf}
    \ketL{\X} \rightarrow
    \ketL{\X}^{\alpha} \equiv e^{i \alpha(\X)} \ketL{\X}
    \;.
\end{align}
While the eigenvalue equation \eqref{ModVec-Eigen} is unaffected by this gauge transformation, the action \eqref{ModVec-WeylActs} of a generic Weyl operator on a modular vector becomes
\begin{align}
\label{ModGau-WeylActs}
    \Wey{\Y} \ketL{\X}^\alpha &=
    e^{i \alpha(\X) - i \alpha(\X + \Y)} \,
    e^{\half i \omega(\Y,\X) / \hbar}
    \ketL{\X + \Y}^\alpha
    \;, \qquad
    \X, \Y \in \PSp
    \;.
\end{align}
Similarly, the gauge transformation changes the quasi-periodicity relation \eqref{ModVec-QuasiPer} to
\begin{align}
\label{ModGau-QuasiPer}
    \ketL{\X + \K}^{\alpha} &=
    e^{i \beta_{\alpha}(\X,\K)} \ketL{\X}^{\alpha}
    \;, \qquad
    \K \in \Lambda \;,\; \X \in \PSp
    \;,
\end{align}
where
\begin{align}
    \beta_{\alpha}(\X,\K) &\equiv
    \alpha(\X + \K) - \alpha(\X)
    + \frac{1}{2\hbar} \, k \cdot \tk
    + \frac{1}{2\hbar} \, \omega(\K,\X)
    \;.
\end{align}
Hence, it changes the condition \eqref{ModWaveF-QuasiPer} accordingly. The $U(1)$-bundle is modified to $\EL^\alpha \rightarrow \TL$ defined by the identification
\begin{align}
    \EL^\alpha : \quad
    \pr{\theta,\X} \sim
    \pr{\theta \, e^{i \beta_{\alpha}(\X,\K)} ,
        \X + \K }
    \;, \quad
    \K \in \Lambda \;,\; \X \in \PSp \;,\; \theta \in U(1)
    \;.
\end{align}
The modular connection $\AAA$ also transforms under this gauge transformation such that
\begin{align}
\label{ModGau-GaugeTransf}
    \AAA_A(\X) \rightarrow
    \AAA_A(\X) + \hbar \, \pd_A\alpha(\X)
    \;.
\end{align}
Note that the curvature $\omega = \dd\AAA$ of the modular connection is invariant under the gauge transformations.

For modular vectors $\ketL{\X}^{\alpha}$ in a generic gauge $\alpha$, we can write the components of the modular connection as
\begin{align}
    \label{Convenient-Connection}
    \AAA_A(\X) = \half \, \X^B \, \omega_{BA} + \hbar \, \pd_A \alpha(\X)
    \;.
\end{align}

While the modular vectors defined in the last section had their gauge fixed as $\alpha = 0$, we will consider an arbitrary choice of gauge hereafter, even though we often omit the label $\alpha$ for better readability. We will also find out in the next section that a specific gauge fixing is required to obtain the Schr\"odinger and momentum representations as singular limits of the modular ones.

\subsection{Singular limits of modular representations}
\label{sec:SingularLimits}

Roughly speaking, the Schr\"odinger and momentum representations correspond to the limits of the family of modular representations when the spacing of the modular lattice goes to infinity and zero. In this section, we will discuss the details of this limiting process.

Consider the 1-parameter family of modular lattices $\Lambda = \ell \Z^d \oplus \tell \Z^d$, where $\ell$ and $\tell$ are length and momentum scales such that $\ell \tell = 2\pi\hbar$. The modular space $\TL$ has the size $\ell^d \times \tell^d$. Recall also that the Heisenberg operators are represented in the modular representations by
\begin{align}
    \pr{ \hat{q}^a,\hat{p}_a } \sim
    \pr{ \half \, x^a - \hbar \, \frac{\pd \alpha(\X)}{\pd \tx_a}
        + i\hbar \, \frac{\pd}{\pd \tx_a}
        \,,\,
        \half \, \tx_a + \hbar \, \frac{\pd \alpha(\X)}{\pd x^a}
        - i\hbar \, \frac{\pd}{\pd x^a} }
    \;.
\end{align}
Now, let's consider the limit $\ell \rightarrow \infty$. As the position part of the modular space $\TL$ grows to infinite size and becomes decompactified, its momentum part shrinks to a point. This has two consequences for the representation of the Heisenberg operators: Firstly, the term $\pd / \pd \tx_a$ drops, since the wave functions cannot depend non-trivially on momentum. Secondly, the terms $\hbar \, \pd \alpha(\X) / \pd \tx_a$ and $\tx^a / 2 + \hbar \, \pd \alpha(\X) / \pd x^a$ must be independent of momentum, otherwise they would become ill-defined in the limit. This implies that $\alpha$ must be of the form $\alpha(\X) = - \frac{1}{2\hbar} \, x \cdot \tx + f(x)$ for the limit $\ell \rightarrow \infty$ to be well-defined. Comparing the representation of the momentum operator to the one in the Schr\"odinger representation, we find that the gauge choice
\begin{align}
\label{ModSing-AlpS}
    \alphaS(\X) = - \frac{1}{2\hbar} \, x \cdot \tx + \mathrm{const.}
    \;,
\end{align}
is needed to obtain the Schr\"odinger representation, in which $(\hat{q}^a,\hat{p}_a) \sim (x^a, -i\hbar \, \frac{\pd}{\pd x^a})$. We name \eqref{ModSing-AlpS} the \emph{Schr\"odinger gauge}. One can check with this gauge fixing that \eqref{ModGau-WeylActs} also resembles the action of Weyl operators on Schr\"odinger eigenvectors given by $\Wey{\Y} \ketS{x} = e^{\half i y \cdot \ty / \hbar} \, e^{i x \cdot \ty / \hbar} \ketS{x+y}$.

Our argument for \eqref{ModSing-AlpS} is also supported by the quasi-periodicity phase function $\beta_{\alphaS}(\X,\K) = - k \cdot \tx / \hbar$. Note that this is independent of the momentum winding number $\tk$ as it should be, since the momentum part of the modular space shrinks to a point and any dependency on the momentum winding number would result in an ill-defined phase. A winding number $k$ in the position directions, on the other hand, becomes irrelevant as the configuration space is decompactified.

In the limit $\ell \rightarrow \infty$, the modular lattice transitions to the momentum space. This transition can be understood in a coarse-graining approximation to the momentum space, although it is in fact a singular transition from a discrete set in $2d$ dimensions to a continuous set in $d$ dimensions. The continuous momentum space is qualified as a modular lattice by definition, since it is a maximal subset $\Lambda \subseteq \PSp$ satisfying $\omega(\Lambda,\Lambda) \subset 2\pi\hbar \, \Z$, although in fact $\omega(\Lambda,\Lambda) = \cur{0}$. The modular space $\TL$ also changes its topology as it becomes the Schr\"odinger configuration space.

In order to see how the modular vectors behave in the Schr\"odinger limit, one can expand them in terms of momentum eigenvectors as in the footnote \ref{Footnote-ZakMom}. We find
\begin{align}
\label{ModSing-Vec}
    \lim_{\ell \rightarrow \infty}
    \big( \! \det\tambda \big)^{1/2} \, \ketL{\X}^{\alphaS}
    &=
    \lim_{\tell \rightarrow 0}
    \pr{2\pi\hbar}^{-d/2}
    \big( \! \det\tambda \big)
    \sum_{\tn \in \Z^d} e^{-i x \cdot (\tx + \tambda \tn) / \hbar}
    \, \keto{\tx + \tambda \tn}_{\mathrm{mom}}^{\phantom{\dagger}}
    \nonumber \\ &=
    \pr{2\pi\hbar}^{-d/2}
    \int_{\R^d} \dd^d\tx \;
    e^{-i x \cdot (\tx + \tambda \tn) / \hbar}
    \, \keto{\tx + \tambda \tn}_{\mathrm{mom}}^{\phantom{\dagger}}
    \nonumber \\ &=
    \ketS{x}
    \;.
\end{align}
Hence, up to a normalization factor, the modular vectors converge to the position eigenvectors. This concludes our analysis: Although the limit $\ell \rightarrow \infty$ is a singular one in which the topology of the (modular) configuration space changes, we have enough evidence to identify the Schr\"odinger representation with this limit of modular representations.

A similar discussion applies to the momentum representation in the limit $\ell \rightarrow 0$. However, this limit requires a different choice of gauge fixing, namely
\begin{align}
    \alphaM(\X) = + \frac{1}{2\hbar} \, x \cdot \tx + \mathrm{const.}
    \;.
\end{align}
To the best of our knowledge, this is the first study in literature that addresses the role of gauge fixing for the singular limits and the fact that Schr\"odinger and momentum representations require two different choices of modular gauge.

\section{Path integral construction}
\label{sec:PathIntegralConstruction}

In the previous section, we introduced the mathematical details underlying the modular representations of the Weyl algebra and their relationship with the Schr\"odinger representation. We can finally use these modular representations to construct a path integral and compare this path integral to Feynman's original path integral in the Schr\"odinger representation. This will be the goal of this section. We focus here on the special example of a quantum harmonic oscillator for its simplicity, since it is possible to evaluate Gaussian integrals analytically.

We consider the Hamiltonian operator for a non-relativistic quantum harmonic oscillator, given by
\begin{align}
\label{HamOp1}
    \HOp &=
    \frac{1}{2m} \, g^{-1}(\hat{p},\hat{p})
    + \half \, m \;\! \Omega^2 \;\! g(\hat{q},\hat{q})
    \;,
\end{align}
where $g$ is the flat Euclidean metric on $\R^d$, $m$ is the particle mass, and $\Omega$ is the angular frequency of the oscillator. It will be more convenient to express this Hamiltonian in terms of a positive definite metric $\HG$ on $\PSp$ such that
\begin{align}
    \HOp = \half \, \Omega \, \HG(\QQ,\QQ)
    \;, \qquad
    \HG_{AB} \equiv
    \mtrx{m \Omega \, g_{ab} & 0 \\
        0 & \pr{m\Omega}^{-1} g^{ab} }
    \;.
\end{align}
Note that $\det G = 1$.

\subsection{Schr\"odinger-Feynman path integral}
\label{sec:SFPI}

In this section, we will list some key results from Feynman's path integral in the Schr\"odinger representation. These are well-known in the literature, but they will be useful later as a reference when we compare them to our new path integral in the modular representation.

The transition amplitude between two position eigenvectors over a finite time interval $[t_0,t_f]$ can be expressed via the path integral
\begin{align}
\label{SFPI-PathIntegral}
    {}^{\phantom{\dagger}}_{\mathrm{Sch}}\!\!\!\;\bra{x_f}
    e^{-i \;\! (t_f - t_0) \HOp / \hbar} \ketS{x_0} &=
    \int_{x(t_0) = x_0}^{x(t_f) = x_f} \DS x \,
    \exp\!\cor{\frac{i}{\hbar} \, S_{\mathrm{Sch}}[x]}
    \;.
\end{align}
On the right-hand side, the functional integral runs over all paths from $x_0$ to $x_f$ on the configuration space $\R^d$. The path measure $\DS x$ is defined as
\begin{align}
    \DS x &\equiv
    \lim_{N \rightarrow \infty}
    \pr{ \pr{\frac{-i}{2\pi\hbar} \,
        \frac{m N}{t_f - t_0}}^{d/2} \sqrt{\det g} }^N
    \prod_{n=1}^{N-1} \dd^d x_n
    \;.
\end{align}
The action $S_{\mathrm{Sch}}$ is given by
\subeq{\label{SFPI-Action}\begin{align}
    S_{\mathrm{Sch}}[x] &=
    \int_{t_0}^{t_f} \dd t \, \Lag_{\mathrm{Sch}}(x(t),\dot{x}(t))
    \;, \\
    \Lag_{\mathrm{Sch}}(x,\dot{x}) &=
    \half \, m \;\! g(\dot{x},\dot{x})
    - \half \, m \;\! \Omega^2 \;\! g(x,x)
    \;,
\end{align}}
where the dot $\dot{}$ over a variable denotes its time derivative. We can make a Legendre transformation on the Lagrangian $\Lag_{\mathrm{Sch}}$ to recover the classical Hamiltonian function $\Ham_{\mathrm{Sch}}$, given by
\begin{align}
    \Ham_{\mathrm{Sch}}(x,\tx) &=
    \frac{1}{2m} \, g^{-1}(\tx,\tx)
    + \half \, m \;\! \Omega^2 \;\! g(x,x)
    = \half \, \Omega \, \HG((x,\tx),(x,\tx))
    \;,
\end{align}
where $\tx = \pd \Lag_{\mathrm{Sch}} / \pd \dot{x}$. Defining $\X^A \equiv (x^a,\tx_a)$, we can write the (classical) Hamilton equations as
\begin{align}
\label{SFPI-HamEq}
    \dot{\X}(t) =
    \Omega \, \omega^{-1} \HG \, \X(t)
    \;.
\end{align}

\subsection{Modular path integral}

In this section, we will construct, step by step, a path integral formulation for the transition amplitude ${}^{\alpha}_\Lambda \hspace{-2.5pt} \bra{\X_f} e^{-i \;\! (t_f - t_0) \HOp / \hbar} \ketL{\X_0}^\alpha$ between two modular vectors over a finite time interval $[t_0,t_f]$. We assume here that the gauge $\alpha$ is arbitrary, and that the modular lattice is of the form $\Lambda = \lambda \Z^d \oplus \tambda \Z^d$, where $\lambda$ and $\tambda$ are diagonal $d \times d$-matrices that satisfy $\lambda^c{}_a \tambda_c{}^b = 2\pi\hbar \, \delta_a^b$, as mentioned previously in Section \ref{sec:ModRepr}. We will often omit the labels $\Lambda$ and $\alpha$ on the modular vectors. The Hamiltonian operator is that of a quantum harmonic oscillator given in \eqref{HamOp1}.

\subsubsection*{Decomposition of paths}

Following the idea in Feynman's original derivation \cite{Feynman:1948ur}, we pick a large integer $N \in \N$ and split the interval $[t_0,t_f]$ into $N$ equal pieces $[t_n, t_n + \delta t]$, $n = 0, ..., N-1$, where
\begin{align}
    \delta t &\equiv \frac{t_f - t_0}{N} = t_{n+1} - t_n
    \;, \qquad
    t_n \equiv t_0 + n \, \delta t
    \;, \qquad
    t_N \equiv t_f
    \;.
\end{align}
We decompose the unitary evolution operator into a product of $N$ operators, such that $e^{-i \;\! (t_f - t_0) \HOp / \hbar} = e^{-i \;\! \delta t \;\! \HOp / \hbar} \,\cdots\, e^{-i \;\! \delta t \;\! \HOp / \hbar}$. Next, we insert the resolution of the identity \eqref{ModVec-Completeness} before each of these $N$ unitary operators,
\begin{align}
\label{MPI-DOP-Insert}
    \bra{\X_f} e^{-i \;\! (t_f - t_0) \HOp / \hbar} \ket{\X_0}
    &=
    \bra{\X_f} \pr{\int_{\TL} \dd^{2d}\X_N \ketbra{\X_N}{\X_N}} 
    e^{-i \;\! \delta t \;\! \HOp / \hbar} \,\cdots
    \nonumber \\ &\hspace{1.5cm}
    \cdots\, e^{-i \;\! \delta t \;\! \HOp / \hbar}
    \pr{\int_{\TL} \dd^{2d}\X_1 \ketbra{\X_1}{\X_1}}
    e^{-i \;\! \delta t \;\! \HOp / \hbar} \ket{\X_0}
    \nonumber \\ &=
    \int_\TL \dd^{2d}\X_N \cdots \dd^{2d}\X_1 \braket{\X_f}{\X_N}
    \prod_{n=0}^{N-1} \bra{\X_{n+1}}
    e^{-i \;\! \delta t \;\! \HOp / \hbar} \ket{\X_n}
    \;.
\end{align}
Each of the integrals in \eqref{MPI-DOP-Insert} are over the modular space $\TL$, which means that they are over arbitrary modular cells in the phase space. We are free to specify their integration domains as any modular cell. Since we are going to identify the variables $\X_n$ later as points on a continuous path in $\PSp$, we make the choice that each integral over $\X_n$ (for $n = 1, ..., N$) is taken over $\ML(\X_{n-1}) \subset \PSp$, which is a box-shaped modular cell centered at the previous point $\X_{n-1}$. Hence, we write
\begin{align}
    \int_\TL \dd^{2d}\X_N \cdots \dd^{2d}\X_1 &=
    \int_{\ML(\X_0)} \dd^{2d}\X_1
    \int_{\ML(\X_1)} \dd^{2d}\X_2
    \,\cdots
    \int_{\ML(\X_{N-1})} \dd^{2d}\X_N
    \;.
\end{align}
We can simplify this expression by changing the integration variables. We define
\begin{align}
\label{MPI-DOP-StepVariables}
    \X_n \equiv \X_0 + \sum_{j=0}^{n-1} \delta\X_j
\end{align}
for $n = 1, ..., N$, and we change the integration variables from $\X_n \in \ML(\X_{n-1})$ to $\delta\X_{n-1} \in \ML(0)$, where $\ML(0) = \lambda \left[ -\half , \half \right)^d \oplus \tambda \left[ -\half , \half \right)^d \subset \PSp$ is a box-shaped modular cell centered at the origin. Then, we get
\begin{align}
\label{MPI-DOP-NewExpr}
    \bra{\X_f} e^{-i \;\! (t_f - t_0) \HOp / \hbar} \ket{\X_0}
    &=
    \int_{\ML(0)} \dd^{2d}\delta\X_0 \,\cdots
    \int_{\ML(0)} \dd^{2d}\delta\X_{N-1}
    \nonumber \\ &\hspace{0.5cm}
    \times \braket{\X_f}{\X_N}
    \prod_{n=0}^{N-1} \bra{\X_{n+1}}
    e^{-i \;\! \delta t \;\! \HOp / \hbar} \ket{\X_n}
    \;,
\end{align}
together with the definitions \eqref{MPI-DOP-StepVariables}.

\subsubsection*{Infinitesimal transition amplitude}

We focus on calculating the infinitesimal transition amplitudes $\bra{\X_{n+1}} e^{-i \;\! \delta t \;\! \HOp / \hbar} \ket{\X_n}$ in \eqref{MPI-DOP-NewExpr} up to linear order in $\delta t$. Using the Lie-Trotter product formula, we can split the unitary evolution operator as
\begin{align}
\label{MPI-ITA-Trotter}
    e^{-i \;\! \delta t \;\! \HOp / \hbar} &=
    \exp\cor{- \frac{i}{\hbar} \, \delta t \,
    \frac{1}{2m} \, g^{-1}(\hat{p},\hat{p}) }
    \exp\cor{- \frac{i}{\hbar} \, \delta t \,
    \half \, m \;\! \Omega^2 \;\! g(\hat{q},\hat{q}) }
    + \Ord(\delta t^2)
    \;.
\end{align}
We will also expand the modular vectors in terms of Schr\"odinger and momentum eigenvectors, respectively. Namely, we have
\subeq{\begin{align}
    \ket{\X_n} &=
    \big(\! \det\tambda \big)^{-1/2} \, e^{i \alpha(\X_n)} \,
    e^{\half i x_n \cdot \tx_n / \hbar}
    \sum_{k \in \lambda \Z^d}
    e^{i k \cdot \tx_n / \hbar}
    \ketS{x_n + k}
    \;, \\
    \bra{\X_{n+1}} &=
    \pr{\det\lambda}^{-1/2} \, e^{-i \alpha(\X_{n+1})} \,
    e^{\half i x_{n+1} \cdot \tx_{n+1} / \hbar}
    \sum_{\tk \in \tambda \Z^d}
    e^{i \tk \cdot x_{n+1} / \hbar}
    \, {\langle \tx_{n+1} + \tk \vert}_{\mathrm{mom}}
    \;.
\end{align}}
Using these expressions and omitting the $\Ord(\delta t^2)$ terms in \eqref{MPI-ITA-Trotter}, we find
\begin{align}
    \bra{\X_{n+1}} e^{-i \;\! \delta t \;\! \HOp / \hbar}
    \ket{\X_n} &=
    \pr{2\pi\hbar}^{-d} e^{-i \alpha(\X_{n+1}) + i \alpha(\X_n)} \,
    e^{\half i \tx_{n+1} \cdot (x_{n+1} - x_n) / \hbar \,
        - \half i x_n \cdot (\tx_{n+1} - \tx_n) / \hbar}
    \nonumber \\ &\hspace{0.5cm} \times
    \sum_{k \in \lambda \Z^d} \sum_{\tk \in \tambda \Z^d}
    e^{- \frac{i}{\hbar} \delta t \, \pr{
        \frac{1}{2m} g^{-1}(\tx_{n+1}+\tk,\tx_{n+1}+\tk)
        + \half m \Omega^2 g(x_n+k,x_n+k)}}
    \nonumber \\ &\hspace{0.5cm} \times
    e^{i \tk \cdot (x_{n+1}-x_n) / \hbar \,
        - i k \cdot (\tx_{n+1}-\tx_n) / \hbar}
    \;.
\end{align}
By defining $\K \equiv (k,\tk) \in \Lambda$ and $\X_n^* \equiv (x_n,\tx_{n+1}) \in \PSp$, we can formulate the last expression more compactly as
\begin{align}
\label{MPI-ITA-CompactSum}
    \bra{\X_{n+1}} e^{-i \;\! \delta t \;\! \HOp / \hbar}
    \ket{\X_n} &=
    \pr{2\pi\hbar}^{-d} e^{-i \alpha(\X_{n+1}) + i \alpha(\X_n)} \,
    e^{\frac{i}{2\hbar} \;\! \omega(\X_n^*,\delta\X_n)} \,
    e^{- \frac{i}{2\hbar} \;\! \Omega \, \delta t \, \HG(\X_n^*,\X_n^*)}
    \nonumber \\ &\hspace{0.5cm} \times
    \sum_{\K \in \Lambda}
    e^{- \frac{i}{2\hbar} \;\! \Omega \, \delta t \, \HG(\K,\K)} \,
    e^{- \frac{i}{\hbar} \;\! \Omega \, \delta t \, \HG(\K,\X_n^*)} \,
    e^{\frac{i}{\hbar} \;\! \omega(\K,\delta\X_n)}
    \;.
\end{align}
It is easier to handle the infinite sum in this expression if we express it in terms of Jacobi's theta function, whose properties are well-studied. Jacobi's theta function (in $2d$ dimensions), $\vartheta : \C^{2d} \times \mathfrak{H}_{2d} \rightarrow \C$, is defined over a complex vector space $\C^{2d}$ and the Siegel upper-half space\footnote{The \emph{Siegel upper-half space} $\mathfrak{H}_{2d}$ is defined as the set of symmetric, complex $2d \times 2d$-matrices whose imaginary parts are positive definite.} $\mathfrak{H}_{2d}$ by
\begin{align}
\label{MPI-ITA-DefTheta}
    \vartheta(z,\tau) \equiv \sum_{n \in \Z^{2d}}
    \exp\cor{i \pi \, n^T \tau \, n + 2\pi i \, n^T z}
    \;.
\end{align}
Some important properties of this function are included in Appendix \ref{app:Theta}.

In our case, we have a sum over the modular lattice $\Lambda = \Hambda \Z^{2d}$, where $\Hambda^A{}_B \equiv \lambda^a{}_b \oplus \tambda_a{}^b$. The matrix $\Xi \equiv - \frac{\Omega \, \delta t}{2\pi\hbar} \, \Hambda^T \HG \, \Hambda$ is however real, and thus not in the Siegel upper-half space $\mathfrak{H}_{2d}$. In order to avoid this problem, we add a small imaginary part to $\Xi$ and consider $\Xi_\epsilon \equiv \Xi + i\epsilon$ instead, where $\epsilon$ is a positive definite matrix. Hence, we can express \eqref{MPI-ITA-CompactSum} as
\begin{align}
\label{MPI-ITA-CompactTheta}
    \bra{\X_{n+1}} e^{-i \;\! \delta t \;\! \HOp / \hbar}
    \ket{\X_n} &=
    \pr{2\pi\hbar}^{-d} e^{-i \alpha(\X_{n+1}) + i \alpha(\X_n)} \,
    e^{\frac{i}{2\hbar} \;\! \omega(\X_n^*,\delta\X_n)} \,
    e^{- \frac{i}{2\hbar} \;\! \Omega \, \delta t \, \HG(\X_n^*,\X_n^*)}
    \nonumber \\ &\hspace{0.5cm} \times
    \vartheta\!\pr{
        \Xi_\epsilon \;\! \Hambda^{-1} \;\! \X^*_n
        + \frac{1}{2\pi\hbar} \, \Hambda^T \omega \, \delta\X_n
        ,\, \Xi_\epsilon}
    \;.
\end{align}
One important feature of Jacobi's theta function is the inversion identity \eqref{Theta-Inversion}, which is included in the Appendix \ref{app:Theta}. Using this identity, we get
\begin{align}
    \vartheta\!\pr{
        \Xi_\epsilon \;\! \Hambda^{-1} \;\! \X^*_n
        + \frac{1}{2\pi\hbar} \, \Hambda^T \omega \, \delta\X_n
        ,\, \Xi_\epsilon}
    &=
    \pr{i \, \Omega \, \delta t}^{-d} \,
    e^{-\frac{i}{\hbar} \;\! \omega(\X^*_n,\delta\X_n)} \,
    e^{\frac{i}{2\hbar} \;\! \Omega \, \delta t \, \HG_\epsilon(\X^*_n,\X^*_n)}
    \nonumber \\ &\hspace{0.5cm} \times
    \vartheta\!\pr{
        \Hambda^{-1} \;\! \X^*_n
        + \frac{1}{2\pi\hbar} \, \Xi_\epsilon^{-1} \;\!
        \Hambda^T \;\! \omega \, \delta\X_n
        ,\, - \Xi_\epsilon^{-1}}
    \nonumber \\ &\hspace{0.5cm} \times
    \exp\!\cor{\frac{i}{2\hbar} \pr{\Omega \, \delta t}^{-1}
        \delta\X_n^T \, \omega^T \HG_\epsilon^{-1} \omega \, \delta\X_n}
    \;.
\end{align}
Inserting this equation back into \eqref{MPI-ITA-CompactTheta} and noting that $\omega^T \HG^{-1} \omega = \HG$, we find
\begin{align}
    \bra{\X_{n+1}} e^{-i \;\! \delta t \;\! \HOp / \hbar}
    \ket{\X_n} &=
    \pr{2\pi i\hbar \, \Omega \, \delta t}^{-d}
    e^{-i \alpha(\X_{n+1}) + i \alpha(\X_n)} \,
    e^{-\frac{i}{2\hbar} \;\! \omega(\X^*_n,\delta\X_n)} \,
    e^{\frac{i}{2\hbar} \frac{1}{\Omega \;\! \delta t} \,
        \HG_\epsilon(\delta\X_n, \delta\X_n)}
    \nonumber \\ &\hspace{0.5cm} \times
    \vartheta\!\pr{
        \Hambda^{-1} \;\! \X^*_n
        + \frac{1}{2\pi\hbar} \, \Xi_\epsilon^{-1} \;\!
        \Hambda^T \;\! \omega \, \delta\X_n
        , - \Xi_\epsilon^{-1}}
    \;.
\end{align}
Finally, we can use this expression to write the transition amplitude in \eqref{MPI-DOP-NewExpr} as
\begin{align}
\label{MPI-DOP-Final}
    \bra{\X_f} e^{-i \;\! (t_f - t_0) \HOp / \hbar} \ket{\X_0}
    &=
    \int_{\ML(0)} \dd^{2d}\delta\X_0 \,\cdots
    \int_{\ML(0)} \dd^{2d}\delta\X_{N-1}
    \pr{2\pi i\hbar \, \Omega \, \delta t}^{-Nd}
    \braket{\X_f}{\X_N}
    \nonumber \\ &\hspace{0.5cm} \times 
    \prod_{n=0}^{N-1} \bigg(
    e^{-i \alpha(\X_{n+1}) + i \alpha(\X_n)} \,
    e^{-\frac{i}{2\hbar} \;\! \omega(\X^*_n,\delta\X_n)} \,
    e^{\frac{i}{2\hbar} \frac{1}{\Omega \;\! \delta t} \,
        \HG_\epsilon(\delta\X_n, \delta\X_n)}
    \nonumber \\ &\hspace{2.0cm} \times
    \vartheta\!\pr{
        \Hambda^{-1} \;\! \X^*_n
        + \frac{1}{2\pi\hbar} \, \Xi_\epsilon^{-1} \;\!
        \Hambda^T \;\! \omega \, \delta\X_n
        , - \Xi_\epsilon^{-1}}
    \!\bigg)
    \;.
\end{align}

\subsubsection*{Limit $N \rightarrow \infty$}

In order to reformulate \eqref{MPI-DOP-Final} as a path integral, we need to take the limit $N \rightarrow \infty$, or equivalently $\delta t \rightarrow 0$. For this limit to be well-defined, we need to hold the ratio
\begin{align}
    \dX_n \equiv \frac{\delta \X_n}{\delta t}
\end{align}
fixed during the limiting process. The variable $\dX_n$ will be interpreted as the velocity of a path $\X : [t_0,t_f] \rightarrow \PSp$ at the time $t_n$.

This limit has several consequences for the expression \eqref{MPI-DOP-Final}. Firstly, we can make a Taylor expansion around $\delta t = 0$ to find
\begin{align}
    -i \alpha(\X_{n+1}) + i \alpha(\X_n)
    - \frac{i}{2\hbar} \, \omega(\X^*_n,\delta\X_n)
    &=  - \frac{i}{\hbar} \, \delta t \, \dX_n^A \, \AAA_A(\X_n)
    + \Ord(\delta t^2)
    \;.
\end{align}
Secondly, since $\Xi_\epsilon^{-1} \propto \delta t^{-1}$, the theta function in \eqref{MPI-DOP-Final} converges to $1$ as $\delta t \rightarrow 0$ due to the property \eqref{Theta-Limit} of Jacobi's theta function, which is included in the Appendix \ref{app:Theta}. Finally, we change the integration variables once again from $\delta \X_n$ to $\dX_n$. Hence, up to terms of order $\Ord(\delta t^2)$, we get
\begin{align}
\label{MPI-Lim-Dawn}
    \bra{\X_f} e^{-i \;\! (t_f - t_0) \HOp / \hbar} \ket{\X_0}
    &=
    \int_{\frac{1}{\delta t} \ML(0)} \dd^{2d}\dX_0 \,\cdots
    \int_{\frac{1}{\delta t} \ML(0)} \dd^{2d}\dX_{N-1}
    \pr{\frac{\delta t}{2\pi i\hbar \, \Omega}}^{Nd}
    \braket{\X_f}{\X_N}
    \nonumber \\ &\hspace{0.5cm} \times
    \prod_{n=0}^{N-1}
    \exp\!\cor{\frac{i}{\hbar} \, \delta t \pr{
        - \dX_n^A \, \AAA_A(\X_n)
        + \frac{1}{2\Omega} \, \HG(\dX_n, \dX_n) }}
    \;,
\end{align}
where we dropped the $i\epsilon$ scheme as it is not needed any more.
The inner product $\braket{\X_f}{\X_N}$ in this expression can be evaluated using \eqref{ModVec-Inner} as
\begin{align}
    \braket{\X_f}{\X_N} &=
    \sum_{\WW \in \Lambda}
    e^{i \alpha(\X_f + \WW) - i \alpha(\X_f)} \,
    e^{\half i w \cdot \tilde{w} / \hbar} \,
    e^{\half i \omega(\WW,\X_f) / \hbar} \,
    \delta^{2d}(\X_f + \WW - \X_N)
    \;.
\end{align}
The new parameter $\WW \equiv (w,\tilde{w}) \in \Lambda$ that enters the modular path integral here will soon play an important role.

We can finally take the limit $N \rightarrow \infty$ and write \eqref{MPI-Lim-Dawn} as a path integral in $\PSp$. We introduce the path function $\X : [t_0,t_f] \rightarrow \PSp$ as
\begin{align}
    \X(t_n) \equiv \X_n
    = \X_0 + \delta t \sum_{j=0}^{n-1} \dX_j
    \;.
\end{align}
The Dirac delta term $\delta^{2d}(\X_f + \WW - \X_N)$ restricts the endpoint of these paths to $\X_N = \X_f + \WW$. In the space of all paths in $\PSp$ from $\X_0$ to $\X_f + \WW$, we define the modular path measure
\begin{align}
\label{MPI-Measure}
    \DS \X &\equiv
    \lim_{N \rightarrow \infty}
    \pr{\frac{\delta t}{2\pi i\hbar \, \Omega}}^{Nd}
    \delta^{2d}(\X_f + \WW - \X_N)
    \prod_{n=0}^{N-1} \dd^{2d} \dX_n
    \;.
\end{align}
We also introduce the \emph{modular action}
\subeq{\label{MPI-Action}\begin{align}
    S_{\mathrm{mod}}[\X] &\equiv
    \int_{t_0}^{t_f} \dd t \, \Lag_{\mathrm{mod}}(\X(t),\dX(t))
    \;, \\
\label{MPI-Lagrangian}
    \Lag_{\mathrm{mod}}(\X,\dX) &\equiv
    - \dX \cdot \AAA(\X)
    + \frac{1}{2\Omega} \, \HG(\dX,\dX)
    \;,
\end{align}}
where $\dX(t) \equiv \frac{\dd}{\dd t} \X(t)$ is the velocity function. Moreover, we write for simplicity
\begin{align}
    \beta_{\alpha}(\X_f,\WW) &=
    \alpha(\X_f + \WW) - \alpha(\X_f)
    + \frac{1}{2\hbar} \, w \cdot \tilde{w}
    + \frac{1}{2\hbar} \, \omega(\WW,\X_f)
    \;.
\end{align}
Combining all of these definitions, we are finally able to express the transition amplitude between two modular vectors by the path integral
\begin{align}
\label{MPI-Final}
\boxed{
    \bra{\X_f} e^{-i \;\! (t_f - t_0) \HOp / \hbar}
    \ket{\X_0}
    =
    \sum_{\WW\in\Lambda} e^{i \beta_{\alpha}(\X_f,\WW)}
    \int_{\X(t_0) = \X_0}^{\X(t_f) = \X_f + \WW} \DS\X \,
    \exp\!\cor{\frac{i}{\hbar} \, S_{\mathrm{mod}}[\X]}
}
    \;.
\end{align}
This \emph{modular path integral} is the main result of our paper. It is clearly different from Feynman's path integral \eqref{SFPI-PathIntegral} as their domains consist of trajectories on two different spaces with a different dimensionality. Moreover, the modular path integral displays at least three new features:
\begin{enumerate}
    \item The expression \eqref{MPI-Final} contains a sum over the modular lattice, which is due to the topology of the modular space. The parameter $\WW \in \Lambda$ should be interpreted as a \textbf{winding number} for each path around the modular space.
    \item The paths of each winding number $\WW$ around the modular space obtain an additional phase $\beta_{\alpha}(\X_f,\WW)$ depending on their winding number. This phase can be interpreted as analogous to the \textbf{Aharonov-Bohm phase}.
    \item The \textbf{modular action} \eqref{MPI-Action} is different from the usual action \eqref{SFPI-Action}, especially through its dependence on the time derivatives of both position $x$ and momentum $\tilde{x}$ variables. As we will see in Section \eqref{sec:Analysis-Canonical}, this signifies a larger \emph{modular phase space} with twice the number of dimensions.
\end{enumerate}
We will discuss these points and their implications in the following section. We supplement the expression \eqref{MPI-Final} in Appendix \ref{app:Cons} with the proof of its consistency under modular lattice translations and gauge transformations.

\section{Analysis of the modular action}
\label{sec:Analysis}

In this section, we aim to analyse the new modular action \eqref{MPI-Action} and compare it to the standard Schr\"odinger action \eqref{SFPI-Action}.

\subsection{Stationary paths}

The variation of the modular action \eqref{MPI-Action} with respect to the path $\X$ is given by
\begin{align}
\label{Stat-Variation}
    \delta S_{\mathrm{mod}} &=
    \int_{t_0}^{t_f} \dd t \pr{
    \frac{\dd}{\dd t} \pr{
        - \AAA \cdot \delta\X + \frac{1}{\Omega} \, \HG(\dX,\delta\X)
    } - \delta\X^A \pr{
        \omega_{AB} \, \dX^B + \frac{1}{\Omega} \, \HG_{AB} \, \ddX^B
    }}
\;.
\end{align}
The Euler-Lagrange equation of motion can be read from the (second) bulk term. Using $\HG^{-1} \omega = - \omega^{-1} \HG$, we write it as
\begin{align}
\label{Stat-ModEOM}
    \ddX(t) = \Omega \, \omega^{-1} \HG \, \dX(t)
    \;.
\end{align}
This (Lagrangian) equation of motion is comparable to the Hamilton equations \eqref{SFPI-HamEq} in the Schr\"o\-dinger case, but it contains an additional time derivative. If we integrate \eqref{Stat-ModEOM}, we get
\begin{align}
\label{Stat-ModEOM-Int}
    \dX(t) = \Omega \, \omega^{-1} \HG \pr{\X(t) - \chi}
\end{align}
with a new integration constant $\chi \in \PSp$.

\begin{figure}
    \centering
    \includegraphics[width = 0.42\textwidth]{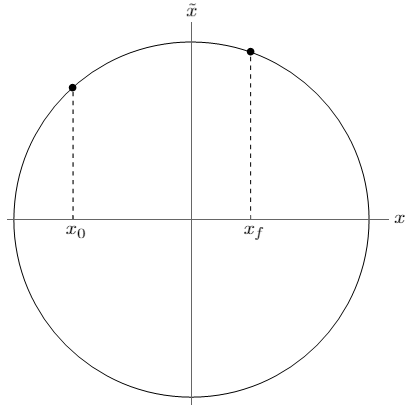}
    \hspace{0.05\textwidth}
    \includegraphics[width = 0.45\textwidth]{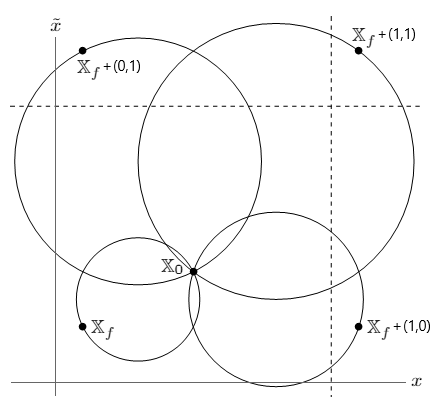}
    \caption{On the left, we have the phase space diagram for a stationary solution to the quantum harmonic oscillator in the Schr\"odinger representation. On the right, four stationary trajectories with different winding numbers in the modular representation are illustrated. These two figures demonstrate the contrast between the trajectories on $\PSp \sim \R^{2d}$ for the two representations.}
    \label{fig:PhaseSpaceDiagrams}
\end{figure}

In order to solve the equation of motion \eqref{Stat-ModEOM}, we note that $\omega^{-1} \HG$ is a complex structure on $\PSp$, i.e. it is a $2d \times 2d$ matrix that satisfies $\pr{\omega^{-1} \HG}^2 = - \one$, where $\one$ is the identity matrix. Combining \eqref{Stat-ModEOM} and \eqref{Stat-ModEOM-Int} gives $\ddX(t) = - \Omega^2 \pr{\X(t) - \chi}$. The solutions to this equation are of the form
\begin{align}
    \X(t) = \chi
    + \xi \sin\!\pr{\Omega t}
    - \omega^{-1} \HG \, \xi \cos\!\pr{\Omega t}
    \;,
\end{align}
where $\xi \in \PSp$ is another integration constant. These integration constants, $\chi$ and $\xi$, are fixed by the boundary conditions of a path $\X(t)$. If we require $\X(t_0) = \X_0$ and $\X(t_f) = \X_f + \WW$ as in the path integral \eqref{MPI-Final}, the stationary paths $\X^{\mathrm{s}}_\WW$ are explicitly given by\footnote{We assume here that $\Omega (t_f - t_0) \notin 2\pi \Z$, since otherwise $\X(t_0) = \X(t_f)$.}
\subeq{\label{Stat-Solution}\begin{align}
    \X^{\mathrm{s}}_\WW(t) &= \chi
    + \xi \sin\!\pr{\Omega\pr{t-\frac{t_f+t_0}{2}}}
    - \omega^{-1} \HG \, \xi \cos\!\pr{\Omega\pr{t-\frac{t_f+t_0}{2}}}
    \;, \\
    \chi &= \half \pr{\X_0 + \X_f + \WW}
    + \half \, \omega^{-1} \HG \pr{\X_f + \WW - \X_0}
        \cot\!\pr{\half \, \Omega \pr{t_f-t_0}}
    \;, \\
    \xi &= \half \pr{\X_f + \WW - \X_0}
        \csc\!\pr{\half \, \Omega \pr{t_f-t_0}}
    \;.
\end{align}}
There are several important distinctions between these stationary paths and the usual result in the Schr\"odinger reprentation.
\begin{itemize}
    \item Firstly, these two sets of paths are defined on different spaces. Schr\"odinger paths run over the corresponding configuration space $\R^d$, whereas the modular paths as in \eqref{Stat-Solution} run over the universal cover of the modular space $\TL$, which is $\R^{2d}$. They are also associated with different phase spaces. The phase space for Schr\"odinger paths is $\PSp = \R^{2d}$, whereas the phase space for modular paths is $\PSp_{\mathrm{mod}} = \R^{4d}$, as we will discuss in Section \ref{sec:Analysis-Canonical}.
    \item The second difference is the number of stationary paths. For any boundary conditions $x(t_0) = x_0$ and $x(t_f) = x_f$, there is a unique classical solution\footnote{Again, we assume $\Omega (t_f - t_0) \notin \pi \Z$, since otherwise $x(t_0) = \pm x(t_f)$.} to the harmonic oscillator in the Schr\"odinger representation. One the other hand, there is one solution \eqref{Stat-Solution} for each winding number $\WW \in \Lambda$ in the modular representation, meaning that there are infinitely many stationary paths in total. This result stems from the compact topology of the modular space.
    \item Heuristically, we can match the phase space $\PSp$ of the Schr\"odinger representation with the universal cover $\R^{2d}$ of the modular space, despite their different physical interpretations. Then, we can compare the solutions in both representations on this common space. The phase space diagram for the Schr\"odinger solution is an ellipse centered at the origin. On the other hand, the paths \eqref{Stat-Solution} are infinitely many ellipses which intersect at the point $\X_0$, see Figure \ref{fig:PhaseSpaceDiagrams}.
\end{itemize}

\subsection{Semi-classical approximation}

For each such stationary path in \eqref{Stat-Solution}, the value of the on-shell modular action is
\begin{align}
\label{SCA-OnShellAction}
    S_{\mathrm{mod}}[\X^{\mathrm{s}}_\WW] &=
    - \hbar \, \alpha(\X_f + \WW) + \hbar \, \alpha(\X_0)
    - \half \, \omega(\X_0, \X_f + \WW)
    \nonumber \\ &\hspace{0.5cm}
    + \frac14 \cot\!\pr{\half \, \Omega \pr{t_f-t_0}}
        \HG(\X_f + \WW - \X_0, \X_f + \WW - \X_0)
    \;.
\end{align}
In the semi-classical approximation $\hbar \rightarrow 0$, the path integral is dominated by the stationary paths. Moreover, since the second functional derivative of the modular action is independent of the winding number, as in
\begin{align}
    \frac{\delta^2 S_{\mathrm{mod}}}{\delta \X^A(t) \, \delta \X^B(t')} &=
    - \, \omega_{AB} \, \frac{\dd}{\dd t} \, \delta(t-t')
    - \frac{1}{\Omega} \, \HG_{AB} \, \frac{\dd^2}{\dd t^2} \, \delta(t-t')
    \;,
\end{align}
each stationary path contributes to the path integral with equal weight. Hence, we find that in the semi-classical approximation the transition amplitude becomes
\begin{align}
\label{SCA-Eq1}
    \bra{\X_f} e^{-i \;\! (t_f - t_0) \HOp / \hbar}
    \ket{\X_0}
    &\underset{\hbar \rightarrow 0}{\sim}
    \sum_{\WW\in\Lambda} e^{i \beta_{\alpha}(\X_f,\WW)} \,
    e^{\frac{i}{\hbar} S_{\mathrm{mod}}[\X^{\mathrm{s}}_\WW]}
    \nonumber \\ &\hspace{4pt}
    = e^{- i \alpha(\X_f) + i \alpha(\X_0)} \,
    e^{- \frac{i}{2\hbar} \, \omega(\X_0, \X_f)}
    \sum_{\WW\in\Lambda}
    e^{\frac{i}{2\hbar} \, w \cdot \tilde{w}} \,
    e^{\frac{i}{2\hbar} \, \omega(\WW,\X_0 + \X_f)} \,
    \nonumber \\ &\hspace{4pt} \hspace{3cm} \times
    e^{\frac{i}{4\hbar}
        \cot(\frac{\Omega}{2} (t_f-t_0)) \,
        \HG(\WW + \X_f - \X_0, \WW + \X_f - \X_0)}
\end{align}
up to a constant factor. We can rewrite this expression in terms of Jacobi's theta function that is defined in \eqref{MPI-ITA-DefTheta} and discussed in Appendix \ref{app:Theta}. For this purpose, we introduce a new metric\footnote{This $O(d,d)$ metric $\eta$ is commonly introduced in several frameworks that are inspired from the T-duality in string theory, including generalized geometry \cite{Hitchin_2003}, double field theory \cite{Hull:2009mi}, and Born geometry \cite{Freidel:2013zga}.} $\eta$ on $\PSp$ defined as $\eta(\X,\Y) \equiv \tx \cdot y + x \cdot \ty$ for any $\X, \Y \in \PSp$. Then, the last expression can be written as
\begin{align}
    &
    \bra{\X_f} e^{-i \;\! (t_f - t_0) \HOp / \hbar}
    \ket{\X_0}
    \nonumber \\[5pt]
    &\underset{\hbar \rightarrow 0}{\sim}
    e^{- i \alpha(\X_f) + i \alpha(\X_0)} \,
    e^{- \frac{i}{2\hbar} \, \omega(\X_0, \X_f)} \,
    e^{\frac{i}{4\hbar} c \,
        \HG(\X_f - \X_0, \X_f - \X_0)}
    \nonumber \\ &\hspace{4pt} \hspace{0.5cm} \hspace{4pt} \times
    \vartheta\!\pr{
        \frac{1}{4\pi\hbar} \, \Hambda^T \pr{
        \omega \pr{\X_0 + \X_f} + c \;\! \HG \pr{\X_f - \X_0}
        }
        ,\,
        \frac{1}{4\pi\hbar} \, \Hambda^T \pr{\eta + c \;\! \HG} \Hambda
        + i \epsilon
        }
    \;,
\end{align}
where $c \equiv \cot(\frac{\Omega}{2} (t_f-t_0))$. We used here the $i\epsilon$ prescription to make the sum converge, where $\epsilon$ is a positive-definite matrix.

\subsection{Canonical analysis}
\label{sec:Analysis-Canonical}

We introduce conjugate momenta $\PP \in \R^{2d}$ to the coordinates $\X \in \PSp$ with respect to the modular action \eqref{MPI-Action}. These are defined as
\begin{align}
    \PP_A \equiv \frac{\delta S_{\mathrm{mod}}}{\delta \dX^A} =
    - \AAA_A(\X)
    + \frac{1}{\Omega} \, \HG_{AB} \, \dX^B
    \;.
\end{align}
The \emph{modular phase space} $\PSp_{\mathrm{mod}} = \R^{4d}$ consists of the pairs of variables $(\X,\PP)$. Note that this has twice the number of dimensions compared to its Schr\"odinger counterpart $\PSp$.

The symplectic potential $\Theta$ on the modular phase space $\PSp_{\mathrm{mod}}$ can be read from the boundary term in the variation of the modular action \eqref{Stat-Variation} as
\begin{align}
    \Theta = \PP_A \, \dd\X^A
    \;.
\end{align}
The exterior derivative of this symplectic potential gives the \emph{modular symplectic form}
\begin{align}
    \omega_{\mathrm{mod}} = \dd\PP_A \wedge \dd\X^A
    \;.
\end{align}
We can also perform a Legendre transform on the modular Lagrangian \eqref{MPI-Lagrangian} to get the \emph{modular Hamiltonian}
\begin{align}
\label{Canon-Hamiltonian}
    \Ham_{\mathrm{mod}}(\X,\PP) =
    \frac{1}{2} \, \Omega \, \HG^{-1}(\PP + \AAA(\X), \PP + \AAA(\X))
    \;.
\end{align}
Hamilton's principal function $S^{\mathrm{s}}_{\mathrm{mod}}(\X,t)$ for this system can be read from the on-shell modular action \eqref{SCA-OnShellAction} as
\begin{align}
\label{Canon-HPFS}
    S^{\mathrm{s}}_{\mathrm{mod}}(\X,t) &=
    - \hbar \, \alpha(\X) + \hbar \, \alpha(\X_0)
    - \half \, \omega(\X_0, \X)
    \nonumber \\ &\hspace{0.5cm}
    + \frac14 \cot\!\pr{\half \, \Omega \pr{t - t_0}}
        \HG(\X - \X_0, \X - \X_0)
    \;.
\end{align}
This function satisfies the Hamilton-Jacobi equation for the Hamiltonian \eqref{Canon-Hamiltonian}.

\subsection{Symmetries}

In this section, we will discuss some of the symmetries of the modular action in \eqref{MPI-Action}. While the symmetries (e.g., rotation and time translation) of the old action \eqref{SFPI-Action} are still present with formally different currents, we find here a whole new set of symmetries that correspond to translations over the phase space, i.e.~spatial translations and momentum translations.

In the following, we give a list of some symmetries of the modular harmonic oscillator together with their corresponding Noether currents.
\begin{itemize}
\item \textbf{Phase space translation}

Spatial and momentum translations are not among the symmetries of the Schr\"odinger harmonic oscillator, but we will show here that they are a new set of symmetries for the modular action \eqref{MPI-Action}. Consider an infinitesimal translation of the phase space coordinates by a constant vector $\mathcal{E} \in \PSp$, such that
\begin{align}
    \delta \X^A &= \mathcal{E}^A
    \;.
\end{align}
The modular Lagrangian changes by a total derivative,
\begin{align}
    \delta \Lag_{\mathrm{mod}} &=
    \frac{\dd}{\dd t} \pr{
    - \mathcal{E} \cdot \AAA(\X)
    + \omega(\X,\mathcal{E}) }
    \;.
\end{align}
We find that the Noether current for this transformation is given by
\begin{align}
    \chi &= \X(t)
    + \frac{1}{\Omega} \, \omega^{-1} \HG \, \dX(t)
    \;,
\end{align}
which is no different than the integration constant we found in \eqref{Stat-ModEOM-Int}. This quantity is conserved on-shell and it denotes the midpoint of the elliptical trajectories we found in \eqref{Stat-Solution}.

The new conserved current $\chi$ vanishes in the Schr\"odinger limit where the classical Hamilton equations \eqref{SFPI-HamEq} are imposed. Therefore, it has no analog in the Schr\"odinger mechanics.

\item \textbf{Time translation}

Consider an infinitesimal shift of the time parameter $t \rightarrow t + \epsilon$, which results in
\begin{align}
    \delta \X^A &= \epsilon \;\! \dX^A
    \;, \qquad
    \delta \Lag_{\mathrm{mod}} =
    \frac{\dd}{\dd t} \pr{\epsilon \;\! \Lag_{\mathrm{mod}}}
    \;.
\end{align}
The associated Noether current is the total energy, given by
\begin{align}
    E = \frac{1}{2\Omega} \, \HG(\dX(t),\dX(t))
    \;.
\end{align}
Although they are formally different, this expression for conserved energy recovers the usual formula $E = \half m g(\dot{x},\dot{x}) + \half m \Omega^2 g(x,x)$ when the classical Hamilton equations \eqref{SFPI-HamEq} are imposed.

\item \textbf{Rotation}

For any infinitesimal, anti-symmetric 2-tensor $L_{ab}$ on $\R^d$, we consider the rotation as
\begin{align}
    \delta x^a = - g^{ab} L_{bc} \, x^c
    \;, \qquad
    \delta \tx_a = - L_{ab} \, g^{bc} \, \tx_c
    \;.
\end{align}
The modular Lagrangian changes by a total derivative,
\begin{align}
    \delta \Lag_{\mathrm{mod}} &=
    \frac{\dd}{\dd t} \pr{
    \AAA_a(\X) \, L^a{}_b \, x^b
    + \AAA^a(\X) \, L_a{}^b \, \tx_b
    - \tx_a \, L^a{}_b \, x^b }
    \;,
\end{align}
where we used the metric $g$ to raise and lower indices on $\R^d$. The conserved Noether current is given by
\begin{align}
    J^{ab} &=
    \tx^{[a} x^{b]}
    - m \, \dot{x}^{[a} x^{b]}
    - \frac{1}{m\Omega^2} \, \dot{\tx}^{[a} \tx^{b]}
    \;.
\end{align}
Once again, although they are formally different, this expression recovers the angular momentum $J^{ab} = x^{[a} \tx^{b]}$ when the classical Hamilton equations \eqref{SFPI-HamEq} are imposed.

\item \textbf{Symplectic transformation}

Finally, we consider an infinitesimal transformation of the form $\delta\X = \epsilon \, \omega^{-1} \HG \, \X$, or equivalently,
\begin{align}
    \delta x^a = \frac{\epsilon}{m\Omega} \, g^{ab} \, \tx_b
    \;, \qquad
    \delta \tx_a = - \epsilon \;\! m \;\! \Omega \, g_{ab} \, x^b
    \;.
\end{align}
The modular Lagrangian changes again by a total derivative,
\begin{align}
    \delta \Lag_{\mathrm{mod}} &=
    \frac{\dd}{\dd t} \pr{
    - \epsilon \, \AAA(\X) \, \omega^{-1} \HG \, \X
    + \frac{\epsilon}{2} \, \HG(\X,\X)
    }
    \;.
\end{align}
We find the conserved Noether current
\begin{align}
    \kappa &=
    \frac{1}{\Omega} \, \omega(\X,\dX)
    - \half \, \HG(\X,\X)
    \;.
\end{align}
This quantity is not independent of the previous conserved currents and it can be written as
\begin{align}
    \kappa = \frac{1}{\Omega} \, E
    - \half \, \HG(\chi,\chi)
    \;.
\end{align}
Note that this symmetry mixes the variables $x$ and $\tx$, therefore it is a hidden symmetry for the Schr\"odinger action. As in this example, the modular action can promote hidden symmetries to explicit symmetries.
\end{itemize}
Looking at the above examples, we can draw the following conclusions:
\begin{enumerate}
    \item The symmetries of the standard action are maintained in the modular action. The corresponding Noether currents can be formally different in the new modular formulation, but they recover their standard expressions under the classical equations of motion.
    \item The modular action has a new set of translation symmetries for both position and momentum variables. The corresponding Noether currents vanish under the classical equations of motion.
    \item Since the modular action is formulated on the classical phase space, the hidden symmetries that mix the configuration variable $x$ with the conjugate momentum $\tx$ can be expressed as explicit symmetries of the action for the composite configuration variable $(x,\tx)$.
\end{enumerate}
We conjecture that these three conclusions hold in general for any modular action, i.e.~any action that is derived in the same way from the modular representation of an arbitrary physical system.

Finally, it is worth mentioning that the modular action \eqref{MPI-Action} is also invariant under the $U(1)$ gauge symmetry in \eqref{ModGau-GaugeTransf}. When the modular connection is transformed as $\AAA_A \rightarrow \AAA_A + \hbar \, \pd_A \alpha$ for a scalar function $\alpha$, the modular Lagrangian changes by a total derivative,
\begin{align}
    \delta \Lag_{\mathrm{mod}} &=
    \frac{\dd}{\dd t} \pr{
    - \hbar \, \alpha(\X)
    }
    \;.
\end{align}

\section{Schr\"odinger limit of the modular path integral}
\label{sec:LimitSch}

We discussed previously in Section \ref{sec:SingularLimits} that the Schr\"odinger representation of the Weyl algebra can be identified with the limit of the modular representations as the length scale $\ell$ of the modular lattice goes to infinity. This limit is a singular one, in which the topology of the configuration space changes, nevertheless it is well-defined.

In this section, we show a similar result for the modular path integral \eqref{MPI-Final}. Considering the 1-parameter family of modular lattices $\Lambda = \ell \Z^d \oplus \tell \Z^d$, where $\ell$ is a length scale and $\tell \equiv 2\pi\hbar / \ell$ is a momentum scale, we demonstrate how the path integral \eqref{MPI-Final} in modular space can be identified with the Feynman path integral in Schr\"odinger space (see Section \ref{sec:SFPI}) in the limit $\ell \rightarrow \infty$.

As discussed in Section \ref{sec:SingularLimits}, the Schr\"odinger limit $\ell \rightarrow \infty$ can be well-defined only in the Schr\"odinger gauge \eqref{ModSing-AlpS}. Therefore, we fix the modular gauge in this section as such, i.e.
\begin{align}
\label{LimitSch-Gauge}
    \AAA(\X) = \pr{0, -x}
    \;.
\end{align}
In this gauge, we have
\begin{align}
    \beta_{\alphaS}(\X_f,\WW) &=
    - \frac{1}{\hbar} \, w \cdot \tx_f
\end{align}
and
\begin{align}
\label{LimitSch-Action0}
    S_{\mathrm{mod}}[\X] &=
    \int_{t_0}^{t_f} \dd t \pr{
    x(t) \cdot \dot{\tx}(t)
    + \frac{m}{2} \, g(\dot{x}(t),\dot{x}(t))
    + \frac{1}{2m\Omega^2} \, g^{-1}(\dot{\tx}(t),\dot{\tx}(t))
    }
    \;.
\end{align}
We remark that the term $- \dX \cdot \AAA = x \cdot \dot{\tx}$ in the above expression is reminiscent of relative locality \cite{AmelinoCamelia:2011bm}.

We consider the expression
\begin{align}
\label{LimitSch-Expr1}
    \sum_{\!\phantom{\tell} w \in \ell {\Z}^d \phantom{\tell}\!}
    \sum_{\tilde{w} \in \tell \Z^d}
    e^{- \frac{i}{\hbar} w \cdot \tx_f}
    \int_{\X(t_0) = \X_0}^{\X(t_f) = \X_f + \WW} \DS\X \,
    \exp\!\cor{\frac{i}{\hbar} \, S_{\mathrm{mod}}[\X]}
    \;,
\end{align}
where $\WW = (w,\tilde{w})$, $\tell = 2\pi\hbar / \ell$, and $S_{\mathrm{mod}}[\X]$ is given in \eqref{LimitSch-Action0}. As we change the parameter $\ell$ in \eqref{LimitSch-Expr1}, the functional integral is not affected (except for its boundaries), since it is on $\PSp$, which is independent of $\ell$.

In the limit $\ell \rightarrow \infty$ and $\tell \rightarrow 0$, the modular lattice $\Lambda$ converges to the momentum space in a coarse-graining approximation. Note that this is a singular transition from a countable set in $2d$ dimensions to an uncountable set in $d$ dimensions. The sum over $\tilde{w} \in \tell \Z^d$ approaches an integral over $\tilde{w} \in \R^d$. Recall that the Dirac delta term $\delta^{2d}(\X_f + \WW - \X_N)$ inside the modular path measure \eqref{MPI-Measure} restricts both the position and momentum endpoints of the paths. An integral over $\tilde{w} \in \R^d$ cancels with $\delta^{d}(\tx_f + \tilde{w} - \tx_N)$ and sets the momentum endpoints of the paths free. Then, the expression in \eqref{LimitSch-Expr1} approaches
\begin{align}
\label{LimitSch-Expr2}
    \sum_{w \in \ell {\Z}^d}
    e^{- \frac{i}{\hbar} w \cdot \tx_f}
    \int_{x(t_0) = x_0}^{x(t_f) = x_f + w} \DS x \, \int \DS \tx \,
    \exp\!\cor{\frac{i}{\hbar} \, S_{\mathrm{mod}}[\X]}
    \;,
\end{align}
up to a constant factor. Here, $\DS x$ and $\DS \tx$ are the standard path measures on the Schr\"odinger and momentum spaces, respectively.

We note that the action \eqref{LimitSch-Action0} can be written as
\begin{align}
    S_{\mathrm{mod}}[\X] &= S_{\mathrm{Sch}}[x]
    + \frac{1}{2 m \Omega^2} \int_{t_0}^{t_f} \dd t \,
    g^{-1}\!\pr{\dot{\tx}(t) + m \Omega^2 g x(t),
        \dot{\tx}(t) + m \Omega^2 g x(t)}
    \;,
\end{align}
where $S_{\mathrm{Sch}}[x]$ is given in \eqref{SFPI-Action}. The integral
\begin{align}
    \int \DS \tx \, \exp\!\cor{
    \frac{i}{\hbar} \, \frac{1}{2 m \Omega^2} \int_{t_0}^{t_f} \dd t \,
    g^{-1}(\dot{\tx} + m \Omega^2 g x, \dot{\tx} + m \Omega^2 g x)
    }
\end{align}
is equal to an irrelevant constant. Hence, \eqref{LimitSch-Expr2} becomes
\begin{align}
\label{LimitSch-Expr3}
    \sum_{w \in \ell {\Z}^d}
    e^{- \frac{i}{\hbar} w \cdot \tx_f}
    \int_{x(t_0) = x_0}^{x(t_f) = x_f + w} \DS x \,
    \exp\!\cor{\frac{i}{\hbar} \, S_{\mathrm{Sch}}[x]}
    \;,
\end{align}
up to constant factors. Finally, as $\ell \rightarrow \infty$, the winding modes $w$ become unattainable as they require infinite action. Therefore we set $w = 0$ and remove the sum, getting
\begin{align}
\label{LimitSch-Expr4}
    \int_{x(t_0) = x_0}^{x(t_f) = x_f} \DS x \,
    \exp\!\cor{\frac{i}{\hbar} \, S_{\mathrm{Sch}}[x]}
\end{align}
from the modular path integral as $\ell \rightarrow \infty$. Regarding the left-hand side in \eqref{MPI-Final}, we already found in \eqref{ModSing-Vec} that modular vectors converge to the corresponding position eigenvectors in the limit $\ell \rightarrow \infty$. Hence, we conclude that the modular path integral \eqref{MPI-Final} recovers the Schr\"odinger-Feynman path integral \eqref{SFPI-PathIntegral} in this limit.

\section{Modular Legendre transform}

In the previous sections, we discussed in detail the modular path integral formulation for the quantum harmonic oscillator. One particular feature of the path integral is that it transforms the Hamiltonian of a system to a Lagrangian function. The transformation from the Hamiltonian to the Schr\"odinger Lagrangian, which is found in the Feynman path integral, is well-known and formulated in general as the Legendre transform. However, the modular Lagrangian found in the modular path integral is different from the Schr\"odinger Lagrangian, even though it starts from the same Hamiltonian. This result raises the question of whether we can formulate the transformation from the Hamiltonian to the modular Lagrangian in general in a compact form, analogously to the standard Legendre transform. We call this new transformation the \emph{modular Legendre transform}.

We will construct the modular Legendre transform by following the construction for the harmonic oscillator in Section \ref{sec:PathIntegralConstruction}. In this example, the classical Hamiltonian function was given by $\Ham = \half \, \Omega \, \HG(\Qq,\Qq)$, where $\Qq = (q,p) \in \R^{2d}$.

The first task is to identify the conjugate variables. In the standard Legendre transform, these are the configuration variable $q \in \R^d$ and the conjugate momentum $p \in \R^d$. In the modular framework, Aharonov's modular variables $\X \in \TL = \R^{2d} / \Lambda$ replace the configuration variables. We consider the representation\footnote{We will use here the same symbol for the equivalence classes $\X \in \TL$ and their representatives $\X \in \ML$, abusing the notation.} of the variables $\X$ on an arbitrary modular cell $\ML \subset \R^{2d}$. Once a modular lattice $\Lambda$ is chosen, we can split the variable $\Qq \in \R^{2d}$ into two parts,
\begin{align}
    \Qq = \X + \K
    \;,
\end{align}
where $\X \in \ML$ is a periodic variable and $\K \in \Lambda$ is a discrete variable. Hence, we identify $\X$ as the configuration variable and $\K$ as the conjugate variable in the modular framework. Then, the classical Hamiltonian function for the harmonic oscillator can be written as
\begin{align}
    \Ham(\X,\K) = \frac{\Omega}{2} \, \HG(\X+\K,\X+\K)
    \;.
\end{align}
With an inspiration from the standard Legendre transform, we make the ansatz that the modular Legendre transform can be written in the form
\begin{align}
    \Lag(\X,\dX) = \mathcal{B}(\X,\dX,\K(\X,\dX))
    - \Ham(\X,\K(\X,\dX))
    \;,
\end{align}
where $\mathcal{B}$ is the Berry phase and the function $\K(\X,\dX)$ is to be determined.

The second task is to find the Berry phase. For this purpose, we analyze the step \eqref{MPI-ITA-CompactSum} of the construction in Section \ref{sec:PathIntegralConstruction}. We identify the summation parameter $\K$ in the said equation with our conjugate variable $\K$ here, since they both represent the remainder part in $\Qq$. Moreover, we identify $\frac{i}{\hbar} \, \delta t \, \Lag$ with the exponent in the right-hand side of \eqref{MPI-ITA-CompactSum} in the limit $\delta t \rightarrow 0$. We get
\begin{align}
\label{Legendre-KeyStepExample}
    \Lag = - \dX \cdot \AAA(\X)
    + \omega(\X+\K,\dX)
    - \frac{\Omega}{2} \, \HG(\X+\K,\X+\K)
    \;.
\end{align}
Hence, we find that the Berry phase is given by
\begin{align}
    \mathcal{B} = - \dX \cdot \AAA(\X)
    + \omega(\X+\K,\dX)
    \;.
\end{align}
Note that this Berry phase recovers the standard expression $\mathcal{B} = \tx \cdot \dot{x}$ if we use the Schr\"odinger gauge fixing as in \eqref{LimitSch-Gauge} and set $\K = 0$.

The final task is to determine the function $\dX(\X,\K)$, which shall give $\K(\X,\dX)$ upon inversion. Recall that this step is given in the standard Legendre transform by $\dot{q} = \pd \Ham(q,p) / \pd p$. We would like to imitate this formula by taking the derivative of the Hamiltonian function $\Ham(\X,\K)$ with respect to the conjugate variable $\K$. However, $\K$ is a discrete variable and the said derivative is not well-defined.

Recall that the Hamiltonian $\Ham$ is originally a function of $\Qq = \X+\K$. Therefore, the missing derivative with respect to $\K$ can equivalently be expressed as a partial derivative with respect to $\X$. Hence, we postulate
\begin{align}
    \dX^A = (\omega^{-1})^{AB} \, \frac{\pd \Ham(\X,\K)}{\pd \X^B}
    \;.
\end{align}
For the harmonic oscillator, we find $\dX = \Omega \, \omega^{-1} \HG \, (\X + \K)$ and subsequently $\K = - \X - \Omega^{-1} \omega^{-1} \HG \, \dX$. Inserting this expression for $\K(\X,\dX)$ into the Lagrangian \eqref{Legendre-KeyStepExample} gives the modular Lagrangian function that we found in Section \ref{sec:PathIntegralConstruction}. In conclusion, this reconstruction of the modular Legendre transform produces the correct result that we found through the path integral construction of the harmonic oscillator.

We conjecture that the modular Legendre transform that we found here by inspecting the example of the harmonic oscillator holds in general for all systems. To summarize, we found the following prescription for the modular Legendre transform:
\begin{enumerate}
    \item Start from a Hamiltonian function $\Ham(\Qq)$ on the phase space.
    \item Calculate
    \begin{align}
    \label{Legendre-Zebra}
        \dX^A \equiv (\omega^{-1})^{AB} \, \frac{\pd \Ham(\Qq)}{\pd \Qq^B}
        \;.
    \end{align}
    \item Invert the relation $\dX(\Qq)$ found in \eqref{Legendre-Zebra} to obtain $\Qq(\dX)$.
    \item Evaluate the modular Lagrangian by
    \begin{align}
    \label{Legend-GenLag}
    \boxed{
        \Lag_{\mathrm{mod}}(\X,\dX) =
        - \dX \cdot \AAA(\X) + \omega(\Qq(\dX),\dX) - \Ham(\Qq(\dX))
    }
        \;.
    \end{align}
\end{enumerate}
This prescription can be applied to most physical systems in their Hamiltonian formalism to produce a modular Lagrangian function as in \eqref{Legend-GenLag}. We will explore this opportunity in a subsequent paper.

\section{Conclusion}
\label{sec:Conclusion}

In this paper, we gave a detailed presentation for the modular representations of the Weyl algebra. We used this framework to construct a path integral based on a modular representation. Our result is important for both its mathematical novelties and the physical interpretation it carries.

The modular path integral and the modular action found in it for the quantum harmonic oscillator are different from their standard Schr\"odinger counterparts in several ways:
\begin{itemize}
    \item The domain of the modular path integral consists of trajectories on the modular space, which has twice the number of dimensions as the classical configuration space.
    \item The trajectories in the modular path integral are sequences of superposition states in the Schr\"odinger representation. Therefore, they carry a non-classical interpretation of locality.
    \item The modular action maintains all symmetries of the standard action. Although the associated Noether currents are formally different, they recover their standard expressions under the classical equations of motion. In addition, the modular action also reveals the hidden symmetries of the standard action, which contain a mixing of the phase space variables.
    \item The modular action is invariant under translations in the modular space. This new set of translation symmetries are not found in the standard action and the associated Noether currents vanish under the classical equations of motion.
    \item The modular path integral contains a sum over the winding numbers of the paths around the modular space. Moreover, an Aharonov-Bohm phase that depends on the winding numbers multiplies the path integral.
\end{itemize}
We formulated the transformation from a classical Hamiltonian function to a modular Lagrangian in a novel prescription that we called the modular Legendre transform. While this prescription is derived here from the study of the harmonic oscillator, we propose that it can be applied to a variety of physical systems including field theories and gravity. This opportunity can provide new formulations and understanding of our physical theories. This is subject for future research.

$$$$

\section*{Acknowledgements}

I am grateful to Laurent Freidel for inspiring this work through his research and discussions, and to Lee Smolin for supporting me throughout this project. I am also thankful to Marc Geiller, Flaminia Giacomini, Tom\'a\v{s} Gonda, Tom\'as Reis, and Barbara \v{S}oda for fruitful conversations and reviewing the drafts of this paper. Research at Perimeter Institute is supported in part by the Government of Canada through the Department of Innovation, Science and Economic Development Canada, and by the Province of Ontario through the Ministry of Economic Development, Job Creation and Trade. This research was also partly supported by grants from John Templeton Foundation and FQXi.

$$$$


\appendix

\section{Jacobi's theta function}
\label{app:Theta}

We give here a brief introduction for Jacobi's theta function. We refer the reader to \cite{mumford_1983} for proofs and more details.

For $D \in \N$, let $\mathfrak{H}_D$ denote the set of symmetric $D \times D$ complex matrices whose imaginary part is positive definite. $\mathfrak{H}_D$ is an open subset in $\C^{D(D+1)/2}$ called the \emph{Siegel upper-half space}. Jacobi's theta function $\vartheta : \C^D \times \mathfrak{H}_D \rightarrow \C$ is defined as
\begin{align}
    \vartheta(z,\tau) \equiv \sum_{n \in \Z^{2d}}
    \exp(i \pi \, n^T \tau \, n + 2\pi i \, n^T z)
\end{align}
for any $z \in \C^D$ and $\tau \in \mathfrak{H}_D$. Some important properties of this function are listed in the following.
\begin{lem}[Periodicity]
For all $m \in \Z^D$, $z \in \C^D$ and $\tau \in \mathfrak{H}_D$,
\begin{align}
    \vartheta(z+m,\tau) = \vartheta(z,\tau)
    \;.
\end{align}
\end{lem}
\begin{lem}[Quasi-periodicity]
For all $m \in \Z^D$, $z \in \C^D$ and $\tau \in \mathfrak{H}_D$,
\begin{align}
    \vartheta(z + \tau m, \tau) =
    \exp(-i\pi \, m^T \tau \, m - 2\pi i \, m^T z) \,
    \vartheta(z,\tau)
    \;.
\end{align}
\end{lem}
\begin{lem}
For all $A \in GL(D,\Z)$,\footnote{$GL(D,\Z)$ is defined as the group of invertible $D \times D$ matrices with integer entries, whose inverses are also integer matrices.} and for all $z \in \C^D$ and $\tau \in \mathfrak{H}_D$,
\begin{align}
    \vartheta(A^T z, A^T \tau A) = \vartheta(z,\tau)
    \;.
\end{align}
\end{lem}
\begin{lem}
For all integer, even-diagonal\footnote{An even-diagonal matrix $B$ is one for which $n^T B \, n$ is an even integer for all $n \in \Z^D$.} and symmetric $D \times D$ matrices $B$, and for all $z \in \C^D$ and $\tau \in \mathfrak{H}_D$,
\begin{align}
    \vartheta(z, \tau + B) = \vartheta(z,\tau)
    \;.
\end{align}
\end{lem}
\begin{lem}[Inversion identity]
For all $z \in \C^D$ and $\tau \in \mathfrak{H}_D$,
\begin{align}
\label{Theta-Inversion}
    \vartheta(\tau^{-1} z, - \tau^{-1}) &=
    \det\!\cor{-i\tau}^{1/2}
    \exp\!\cor{i\pi z^T \tau^{-1} z}
    \vartheta(z,\tau)
    \;.
\end{align}
\end{lem}
\begin{lem}
The following limit holds for all $z \in \C^D$ and $\tau \in \mathfrak{H}_D$,
\begin{align}
\label{Theta-Limit}
    \lim_{a \rightarrow +\infty} \vartheta(z,a\tau) = 1
    \;,
\end{align}
where $a \in \R_+$. The convergence is stronger than quadratic, i.e.~$\vartheta(z,a\tau) = 1 + \Ord(a^{-2})$.
\end{lem}

\section{Consistency of the modular path integral}
\label{app:Cons}

In this appendix, we will prove various properties of the modular path integral \eqref{MPI-Final}, which are necessary for its consistency. Throughout this section, we will frequently use the fact that the modular action \eqref{MPI-Action} can be written as
\begin{align}
    \label{Convenient-Action}
    S_{\mathrm{mod}}[\X] &=
    - \hbar \, \alpha(\X(t_f))
    + \hbar \, \alpha(\X(t_0))
    + \int_{t_0}^{t_f} \dd t \pr{
    - \half \, \omega(\X,\dX)
    + \frac{1}{2\Omega} \, \HG(\dX,\dX)
    }
    \;,
\end{align}
which follows from \eqref{Convenient-Connection}.

\subsection{Discrete translations}

Here, we will show that the modular path integral \eqref{MPI-Final} is consistent under a discrete translation of its endpoints. $\K \in \Lambda$ denotes an arbitrary lattice point in this section. The proof consists of two parts.

Firstly, we examine a shift in the final point, i.e.
\begin{align*}
    \bra{\X_f + \K} e^{-i \;\! (t_f - t_0) \HOp / \hbar} \ket{\X_0}
    &=
    \sum_{\WW\in\Lambda} e^{i \beta_{\alpha}(\X_f + \K,\WW)}
    \int_{\X(t_0) = \X_0}^{\X(t_f) = \X_f + \K + \WW} \DS\X \,
    \exp\!\cor{\frac{i}{\hbar} \, S_{\mathrm{mod}}[\X]}
    \\
    &=
    \sum_{\WW\in\Lambda} e^{i \beta_{\alpha}(\X_f + \K, \WW - \K)}
    \int_{\X(t_0) = \X_0}^{\X(t_f) = \X_f + \WW} \DS\X \,
    \exp\!\cor{\frac{i}{\hbar} \, S_{\mathrm{mod}}[\X]}
    \;,
\end{align*}
where we redefined the summation variable $\WW$ in the second line. We have
\begin{align*}
    \beta_{\alpha}(\X_f + \K, \WW - \K) &=
    \beta_{\alpha}(\X_f,\WW) - \beta_{\alpha}(\X_f,\K)
    + \frac{1}{\hbar} \pr{k - w} \cdot \tk
    \;.
\end{align*}
Since $e^{\frac{i}{\hbar} \pr{k - w} \cdot \tk} = 1$, we find
\begin{align*}
    \bra{\X_f + \K} e^{-i \;\! (t_f - t_0) \HOp / \hbar} \ket{\X_0}
    = e^{-i \beta_{\alpha}(\X_f,\K)}
    \bra{\X_f} e^{-i \;\! (t_f - t_0) \HOp / \hbar} \ket{\X_0}
    \;.
\end{align*}
This is consistent with the quasi-periodicity \eqref{ModGau-QuasiPer} of the modular vector.

The second part of the proof consists of examining a shift in the initial point, i.e.
\begin{align*}
    \bra{\X_f} e^{-i \;\! (t_f - t_0) \HOp / \hbar} \ket{\X_0 + \K}
    &=
    \sum_{\WW\in\Lambda} e^{i \beta_{\alpha}(\X_f,\WW)}
    \int_{\X(t_0) = \X_0 + \K}^{\X(t_f) = \X_f + \WW} \DS\X \,
    \exp\!\cor{\frac{i}{\hbar} \, S_{\mathrm{mod}}[\X]}
    \;.
\end{align*}
For any path $t \in [t_0,t_f] \mapsto \X(t)$, let $\X + \K$ denote the parallel path shifted by the constant $\K$. We can shift the integration variable in the path integral and write
\begin{align*}
    \bra{\X_f} e^{-i \;\! (t_f - t_0) \HOp / \hbar} \ket{\X_0 + \K}
    &=
    \sum_{\WW\in\Lambda} e^{i \beta_{\alpha}(\X_f,\WW)}
    \int_{\X(t_0) = \X_0}^{\X(t_f) = \X_f + \WW - \K} \DS\X \,
    \exp\!\cor{\frac{i}{\hbar} \, S_{\mathrm{mod}}[\X + \K]}
    \\
    &=
    \sum_{\WW\in\Lambda} e^{i \beta_{\alpha}(\X_f,\WW + \K)}
    \int_{\X(t_0) = \X_0}^{\X(t_f) = \X_f + \WW} \DS\X \,
    \exp\!\cor{\frac{i}{\hbar} \, S_{\mathrm{mod}}[\X + \K]}
    ,
\end{align*}
where we redefined the summation variable $\WW$ in the second line. Using the expression \eqref{Convenient-Action}, we find that the action transforms as
\begin{align*}
    S_{\mathrm{mod}}[\X + \K] &= S_{\mathrm{mod}}[\X]
    - \half \, \omega(\K, \X(t_f) - \X(t_0))
    - \hbar \, \alpha(\X(t_f) + \K)
    + \hbar \, \alpha(\X(t_f))
    \nonumber \\ &\hspace{0.5cm}
    + \hbar \, \alpha(\X(t_0) + \K)
    - \hbar \, \alpha(\X(t_0))
    \\[5pt]
    &= S_{\mathrm{mod}}[\X]
    - \hbar \beta_{\alpha}(\X(t_f),\K)
    + \hbar \beta_{\alpha}(\X(t_0),\K)
    \;.
\end{align*}
Finally, we note that
\begin{align*}
    \beta_{\alpha}(\X_f,\WW + \K)
    - \beta_{\alpha}(\X_f + \WW,\K)
    &=
    \beta_{\alpha}(\X_f,\WW)
    + \frac{1}{\hbar} \, k \cdot \tilde{w}
    \;.
\end{align*}
Since $e^{\frac{i}{\hbar} k \cdot \tilde{w}} = 1$, we obtain
\begin{align*}
    \bra{\X_f} e^{-i \;\! (t_f - t_0) \HOp / \hbar} \ket{\X_0 + \K}
    &=
    e^{i \beta_{\alpha}(\X_0,\K)}
    \bra{\X_f} e^{-i \;\! (t_f - t_0) \HOp / \hbar} \ket{\X_0}
    \;.
\end{align*}
Once again, this is consistent with the quasi-periodicity \eqref{ModGau-QuasiPer} of the modular vector.

\subsection{Gauge transformations}

Here, we will show that the modular path integral \eqref{MPI-Final} transforms covariantly under a gauge transformation $\AAA_A \rightarrow \AAA_A + \hbar \, \pd_A \bar{\alpha}$. The modular action \eqref{Convenient-Action} and the phase factor transform as
\begin{align*}
    S_{\mathrm{mod}}[\X] &\rightarrow
    S_{\mathrm{mod}}[\X]
    - \hbar \;\! \bar{\alpha}(\X_f + \WW)
    + \hbar \;\! \bar{\alpha}(\X_0)
    \\
    \beta_{\alpha}(\X_f,\WW) &\rightarrow
    \beta_{\alpha}(\X_f,\WW)
    + \bar{\alpha}(\X_f + \WW)
    - \bar{\alpha}(\X_f)
    \;.
\end{align*}
Combining these two expressions, we get
\begin{align*}
    \bra{\X_f} e^{-i \;\! (t_f - t_0) \HOp / \hbar} \ket{\X_0}
    &\rightarrow
    e^{-i \bar{\alpha}(\X_f) + i \bar{\alpha}(\X_0)}
    \bra{\X_f} e^{-i \;\! (t_f - t_0) \HOp / \hbar} \ket{\X_0}
    \;,
\end{align*}
which is consistent with \eqref{ModGau-Transf}.

\bibliography{References.bib}
\bibliographystyle{Utphys}

\end{document}